\documentclass[prl,twocolumn,floatfix,nobibnotes,superscriptaddress]{revtex4}
\usepackage{amsmath}
\usepackage{graphicx}
\usepackage{comment}
\usepackage[tight]{subfigure}

\newcommand{\be}{\begin{equation}}
\newcommand{\ee}{\end{equation}}
\newcommand{\new}[1]{\textcolor{black} {#1}}

\usepackage[usenames,dvipsnames]{color}

\usepackage{color}

\begin{document}
\title{Universal distribution of magnetic anisotropy of impurities in ordered and disordered nano-grains}
\author{A. Szilva}
\affiliation{Department of Physics and Astronomy, Division of Materials Theory, Uppsala University,
Box 516, SE-75120, Uppsala, Sweden}
\affiliation{Department of Theoretical Physics, Budapest University of Technology and Economics, Budafoki \'ut 8. H-1111 Budapest, Hungary}
\author{P. Balla}
\affiliation{Department of Theoretical Physics, Budapest University of Technology and Economics, Budafoki \'ut 8. H-1111 Budapest, Hungary}
\affiliation{Institute for Solid State Physics and Optics, Wigner Research Centre for Physics,
Hungarian Academy of Sciences, H-1525 Budapest, P.O.B. 49, Hungary}
\author{O. Eriksson}
\affiliation{Department of Physics and Astronomy, Division of Materials Theory, Uppsala University,
Box 516, SE-75120, Uppsala, Sweden}
\author{G. Zar\'and}
\affiliation{BME-MTA Exotic Quantum Phases 'Lend\"ulet' Group, Institute of Physics, Budapest University of Technology and Economics, H-1521 Budapest, Hungary}
\author{L. Szunyogh}
\affiliation{Department of Theoretical Physics, Budapest University of Technology and Economics, Budafoki \'ut 8. H-1111 Budapest, Hungary}
\affiliation{MTA-BME Condensed Matter Research Group, Budapest University of
Technology and Economics, Budafoki ´ut 8., H-1111 Budapest, Hungary}

\begin{abstract}

\new{
We examine  the distribution of the magnetic anisotropy (MA) experienced by a magnetic impurity embedded in a metallic nano-grain.
As an example of a generic magnetic impurity with partially filled $d$-shell, we study the case of  $d^{1}$ impurities imbedded into ordered and disordered 
 Au nano-grains,  described in terms of a realistic band structure.   Confinement of the electrons induces a magnetic anisotropy that is large, and can be characterized by 5 real parameters, coupling to the quadrupolar moments of the spin. In ordered (spherical) nano-grains, these parameters exhibit symmetrical structures and reflect the symmetry of the underlying lattice,  while for disordered grains they are randomly distributed and, -- for stronger disorder, -- their distribution is found to be characterized by random matrix theory.  As a result, the probability of having small magnetic anisotropies $K_L$ is suppressed below a characteristic scale 
 $\Delta_E$, which we predict to scale  with the number of atoms $N$ as $\Delta_E\sim 1/N^{3/2}$. This  gives rise to  
anomalies in the  specific heat and the susceptibility at temperatures $T\sim \Delta_E$ and produces distinct  structures in the magnetic excitation spectrum of the clusters, that should be possible to detect experimentally.}
\end{abstract}
\pacs{later}

\maketitle

Magnetic thin films and nano-sized objects are essential ingredients  for high-density magnetic recording.  Magnetic nanoparticles, in particular, 
are considered as most likely  building blocks for future permanent magnets~\cite{coey,sell,nature}. Similar to molecular electronics devices \cite{wege} or thin metallic layers \cite{pB1,pB2},  spin-orbit (SO) coupling plays 
an essential role in  nano-particles:  by  restricting the free motion of the magnetic spins and eventually 
freezing them~\cite{pB5}, it enables spins to store magnetic information.   
Understanding the behavior of the magnetic anisotropy  in systems with quantum confinement 
is hence of crucial importance for nanoscale magnetic materials science. 

SO coupling induced magnetic  anisotropy (MA) appears to be surprisingly large in certain nanoscale and mesoscopic structures. 
A sterling example, where  confinement  induced MA provides   explanation for the observation, 
is the suppression of the Kondo effect in thin films and wires of certain dilute magnetic alloys  \cite{pB1}. 
 As  revealed by a series of experiments  on magnetically doped  thin metallic films~\cite{pB1,pB2},
SO coupling combined  with a geometrical confinement of the electrons' motion induces a 'dead layer' in the vicinity of the surface, where the motion of the 
otherwise free spins  is blocked by MA.  The thickness $d$ of this  'dead layer', consistently explained in terms of surface induced spin anisotropy~\cite{pB1,pB2},  depends on the  particular host   material and   dopants used, but it can be unexpectedly large,  in the range  of  $d\sim 100\AA$.

In  confined  geometries, spin-orbit (SO) interaction can induce magnetic anisotropies by two fundamentally  different mechanisms.  
In metallic compounds of heavy elements with strong SO interaction, the geometry 
of the sample is imprinted  into the spin texture of the \emph{conduction electrons'} wave function. This spin texture varies in space 
close to the surface of the sample, and  induces a position dependent magnetic anisotropy for the magnetic dopants. 
The corresponding host-induced magnetic anisotropy (HSO mechanism),  intensively studied in atomic-scale engineering,  is  
presumably at work in  magnetically doped noble metal samples, where it gives rise to a relatively short-ranged confinement-induced magnetic anisotropy close to sample  surfaces~\cite{pB5}. Much stronger and longer-ranged anisotropy can, however, be generated by the local SO coupling at the magnetic dopant's 
$d$ or $f$-level~\cite{pub1,pub2} in case of magnetic impurities with a partially filled $d$ or $f$ shell, respectively~\cite{pub1}. 
Since, in this case, the spin of the magnetic ion is entangled with the orbital structure of localized $f$ and $d$ states, and couples very strongly to Friedel oscillations, leading to 
the emergence of a   strong  MA (LSO mechanism). While the HSO mechanism appears to be too weak to explain the thin film experiments,  
the  stronger and  more  slowly decaying anisotropy induced by the LSO mechanism seems to give a consistent explanation for the experimental observations~\cite{pub1,pub2,doktori}, and appears to be the dominant mechanism for SO coupling induced MA in confined structures.

The surface-induced MA has been  thoroughly studied  in thin films and in the vicinity of surfaces. 
Surprisingly little is known, however,  about the structure and size of confinement induced MA in 
nano-grains. Here we therefore investigate the LSO mechanism in  metallic grains 
and demonstrate that symmetrically 'ordered' nano-grains and nano-grains with random surfaces 
show very different behaviors. In ordered nano-grains,  the MA constants 
exhibit  regular structures reflecting the symmetry of the grain.  Different atomic shells of the grain 
behave very differently from the point of view of magnetic anisotropy, which  displays  Friedel-like shell-to-shell 
oscillations.  Adding atoms to an ordered grain and thereby making its surface disordered, however, 
changes this picture completely: in such 'disordered' grains, the conduction electron's wave function 
becomes chaotic, and the distribution of MA parameters become gradually more and more random.  The MA distribution is then 
found to be  fairly well captured  by random matrix theory, 
and to be almost independent of the magnetic ion's position.

In the present work, we shall demonstrate  these characteristic properties by focusing on the simplest  case of a 
magnetic impurity in a $d^{1}$ configuration 
embedded into an fcc Au nano-grain host of 100-400 atoms.  This model system captures the generic properties of most magnetic 
impurities and hosts, and allows us to study the roles of local and host SO interactions  simultaneously. 
We  construct the nano-grains by placing Au atoms on a regular fcc lattice starting from a central site, and then adding 'shells', 
defined as groups of atoms that transform into each other under the cubic group ($O_{h}$).   
We refer to nano-grains with only filled shells as \emph{ordered} (or spherical) nano-grains, while 
nano-grains with partially filled outmost shells shall be referred to as  \emph{disordered} (or non-spherical) nanoparticles. 
We also define the  \textit{core} of the grain as the group of atoms   having a complete set of first neighbors. 
To describe the electronic structure of Au nano-grains, we use a tight binding model with $spd$ canonical orbitals and incorporate SO 
coupling of the host atoms non-perturbatively. More technical details  can be found in the Supplemental Material~\cite{appendix}.

To investigate the local SO-induced anisotropy, we shall use the approach of Ref.~\cite{pB12}, and account for local correlations on the magnetic impurity
by means of a generalized Anderson model~\cite{anderson}, which we embed into the  Au grain described above. 
Similar to Anderson's model, our impurity Hamiltonian (the so-called  ionic model \cite{hirst})
 contains three terms: the impurity term, the conduction electron, and the
hybridization terms (see Ref.~\cite{appendix}). In the ground state  $d^1$ configuration, by Hund's third rule, 
the strong local SO interaction aligns the angular momentum of the $d$-electron   antiferromagnetically  with its spin, 
thus forming a  $D^{3/2}$ spin $j=3/2$ multiplet.  This multiplet remains degenerate in a perfect cubic environment, and --  in group theoretical terms -- it 
transforms according to the  four-dimensional  $\Gamma_{8}$ double representation of the cubic point group~ \cite{tinkham}.  

Next, we need to embed this impurity into the host.  Following Anderson, 
we  consider hybridization of the deep $D^{3/2}$ multiplet  only with $s$-type host electrons, since these latter dominate 
the  density of states  near  the Fermi-energy (see Ref. \cite{appendix}). Local cubic symmetry  implies, however,  
that only linear combinations of neighboring $s$ orbitals, transforming  as  $j \sim 3/2$ can hybridize with the 
deep $D^{3/2}$  states. 
The proper $|s_{3/2}\rangle,\;|s_{1/2}\rangle,\;|s_{-1/2}\rangle,\;|s_{-3/2}\rangle$ basis set has been constructed in Ref.~\cite{pB12,doktori} and 
 is reproduced in the Supplemental material,  Ref.~\cite{appendix}. Considering then charge fluctuations to the $d^0$ state
  and performing  a Coqblin-Schrieffer transformation  \cite{hewson,pB17}, we finally arrive at the following simple exchange Hamiltonian,
\begin{equation}
\mathcal{H}_{LSO}=J\sum_{m,m^{\prime}}s_{m}^{\dagger}s_{m^{\prime}}|m^{\prime}%
\rangle\langle m|\;. \label{hlso}%
\end{equation}
Here the  $\{|m\rangle\}$  refer to the states $\{\frac{3}{2},\frac
{1}{2},-\frac{1}{2},-\frac{3}{2}\}$ of the impurity, and $s_{m}^{\dagger}$ creates
appropriate host electrons, while $J$ denotes the strength of the effective  exchange coupling (see also Ref.~\cite{appendix}).  

To handle  the exchange interaction $J$, we can use a diagrammatic  approach similar to Ref. \cite{pB12}. 
The dominant contribution to the MA is, however, simply given by the  Hartee term, generating the  effective spin Hamiltonian, 
\be
H^{L}= \sum_{m,m'} K_{mm^\prime}|m^{\prime} \rangle\langle m|
\nonumber
\ee
with the anisotropy matrix  $K_{mm^\prime} $ expressed as
\be
K_{mm^\prime} = J \langle s_{m}^{\dagger}s_{m^{\prime}}\rangle = 
J\int_{-\infty}^{\varepsilon_{F}}d\varepsilon\;
\rho_{mm^{\prime}}^{L}(\varepsilon)\;. \label{rho-int}%
\ee
Here $\rho^{L}(\varepsilon)$ denotes the local spectral function matrix of the  symmetry adapted host operators, $s^\dagger_m$,
and $\varepsilon_{F}$ stands for the Fermi-energy. In practice, we evaluate the integral \eqref{rho-int} in terms of the Green's functions 
of the host \footnote{In the integral defining the MA energy, see Eq. \eqref{rho-int}, a cut-off determined by
  $\Delta_{SO}$ should be, in principle, used. The results presented in the paper are calculated without using a cut-off. We checked however numerically that by using a cut-off beyond  $\Delta_{SO}$ =1.5 -2 eV, the MAE does not show signifact changes, therefore, our results are relevant to the limit of large  $\Delta_{SO}$ (~1.5 eV).}.

\begin{figure*}[ht]
\centering
\begin{center}
\begin{tabular}
[c]{c}%
\includegraphics[width=0.8\textwidth, bb=  -70 0 1720 910] {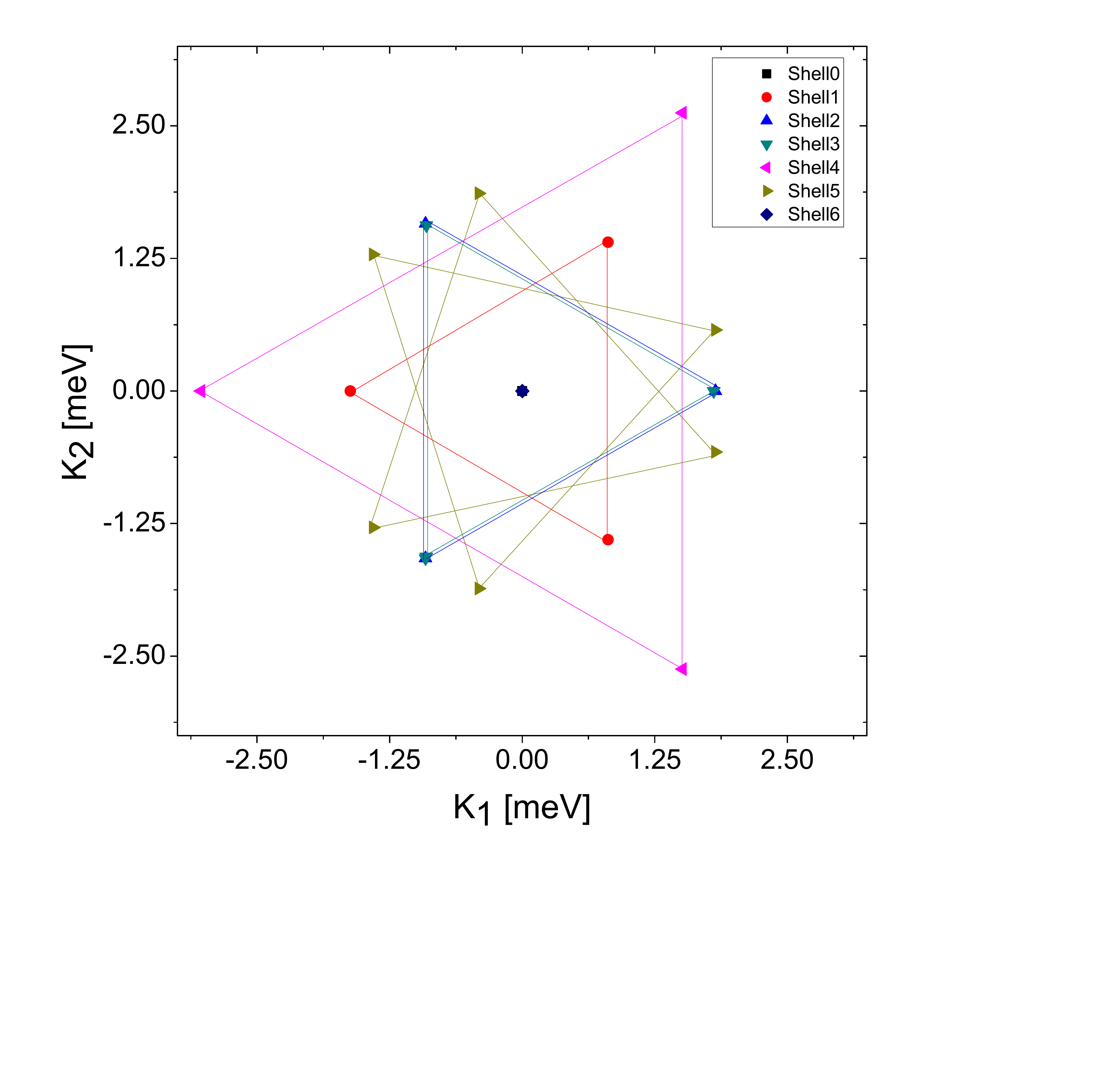} 
\includegraphics[width=0.8\textwidth, bb= 630 0 2300 450] {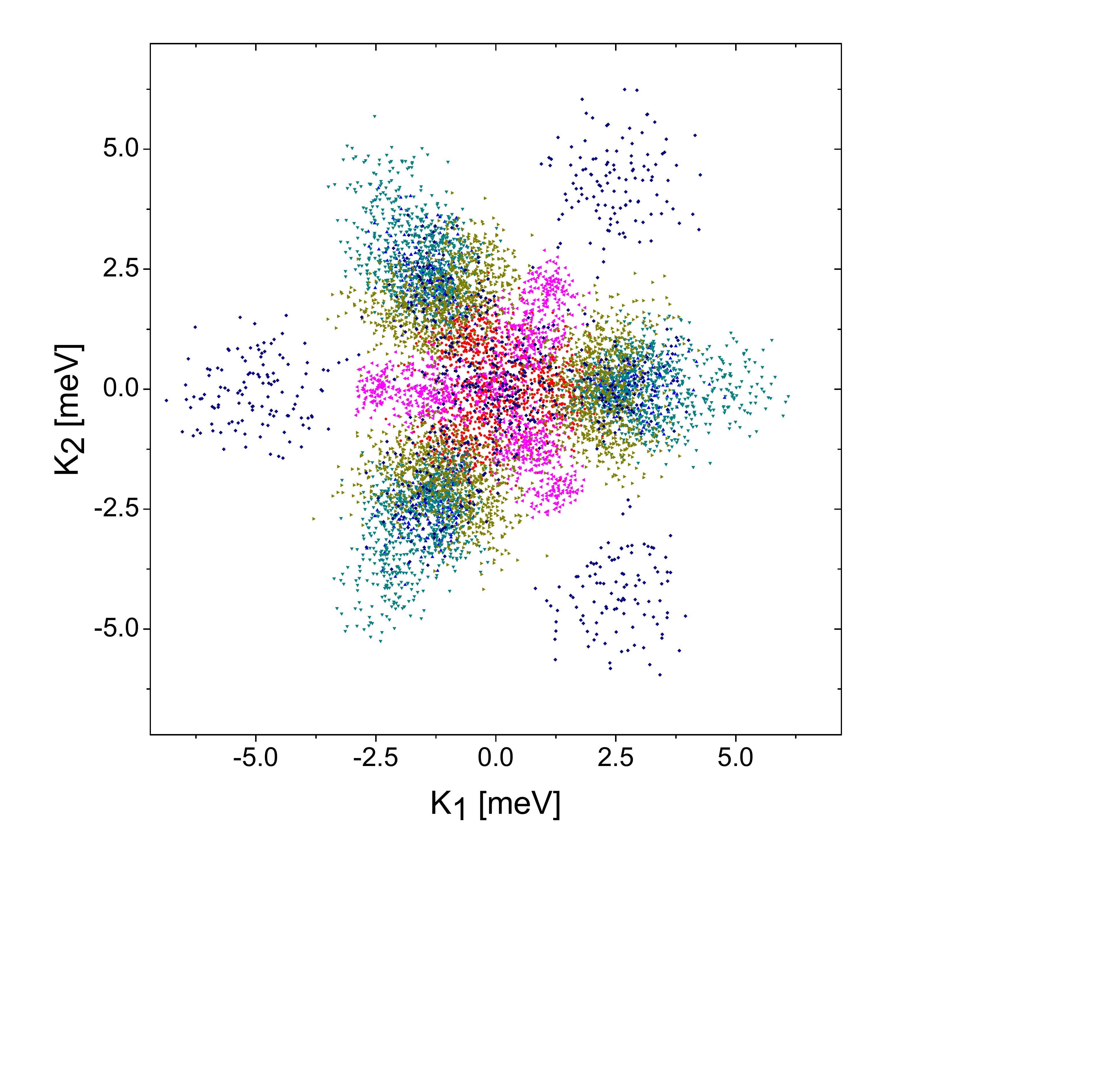}
\end{tabular}
\end{center}
\par
\vskip -0.5cm  \caption{(color online) 
Left: Anisotropy parameters
in the $E$-plane for an ordered grain of $225$ atoms ($87$ core atoms). Right: Anisotropies in the  $E$-plane in case of 100 disordered Au nano-grains. An ordered cluster is created by 225 atoms that fill completely the first  six shells of the fcc structure, and with 11 extra atoms, randomly placed onto the next shell. The triangular structure of the  anisotropy parameter distribution can still be observed.}
\label{tri}%
\end{figure*}

Although disordered nano-grains do not possess  spatial symmetries,  time reversal (TR) symmetry is still 
present, and implies that, apart from an unimportant overall shift $K_0$, the anisotropy matrix  $K_{mm^\prime}$ can be parametrized in terms of five 
real numbers, $K_{\mu}$ ($\mu=1...5$), 
\begin{equation}
\left(
\begin{array}
[c]{cccc}%
K_{1} & K_{3}- K_{5}i & K_{2}- K_{4} i & 0\\
K_{3} + K_{5}i & - K_{1} & 0 & K_{2} - K_{4}i\\
K_{2} + K_{4}i & 0 & - K_{1} & -K_{3} + K_{5}i\\
0 & K_{2} + K_{4}i & - K_{3}- K_{5}i & K_{1}%
\end{array}
\right) \;, \label{aniLSO2}%
\end{equation}
which we shall   refer to as the  LSO-MA parameters.
We note that the absence of SO coupling on the host atoms further simplifies the structure of $H^{L}$, and the 
matrix elements  $K_{3}$, $K_{4}$ and $K_{5}$ vanish in case  the up and down spin channels do not mix in the host.

The MA matrix in Eq. (\ref{aniLSO2}) has  two Kramers degenerate eigenvalues, $\lambda_{+} = -\lambda_-$,
whose splitting  can be used to define naturally the \textit{magnetic anisotropy constant} as
\begin{equation}
K_L\equiv ({\lambda_{+}-\lambda_{-})}/{2}\; 
=  \sqrt{\textstyle \sum_\mu K_\mu^2} \;. \label{K-LSO2}%
\end{equation}
If the magnetic impurity is placed in an ordered (spherical) nano-grain then the MA matrices can be different for different sites even if they belong to the same shell. Their trace, eigenvalues and, therefore, the MA constant should, however, be the same for all sites within the same shell due to the underlying 
($O_{h}$) symmetry of the grain, as indeed confirmed  by our numerical simulations, discussed below (for additional information, see Ref.~\cite{appendix}).

To find  connections between the elements of $K_{mm^\prime}$ for an ordered nano-grain, we express Eq. (\ref{aniLSO2}) in a multipolar basis.
Time reversal symmetry implies that $K_{mm^\prime}$ can be expressed solely in terms of even powers of the $j=3/2$ spin operators, $J_{x},J_{y},J_{z}$. In fact, 
the parameters in \eqref{aniLSO2} couple directly to the usual 5 (normalized and traceless) quadrupole operators allowed by time reversal symmetry, 
$Q_1,\dots,Q_5$ proportional to 
  $ 2J_{z}^{2}-J_{x}^{2}-J_{y}^{2}$, $ J_{x}^{2}-J_{y}^{2}$,   $J_{x}J_{z}+J_{z}J_{x}$, 
    $J_{x}J_{y}+J_{y}J_{x}$ and $J_{y}J_{z}+J_{z}J_{y}$, respectively.
The local Hamiltonian  can  be simply expressed in terms of these as 
\begin{equation}
H^L = \sum_\mu K_\mu Q_\mu\;,
\label{sop}
\end{equation}
with the coefficients $K_\mu$ of the $Q$-matrices forming a 5-dimensional vector. Under cubic point group transformations, the first two components, $\sim (Q_1,Q_2)$
and the last three components, $\sim (Q_3,Q_4,Q_5)$ transform into each other according to the $E$ and $T_{2}$ representations 
of the cubic point group, respectively \cite{tinkham}. 
Correspondingly, for atoms on the same shell of an ordered grain, the anisotropy parameters $K_{1,2}$ transform into each other, and form 
regular patterns of triangular symmetry in the $(K_1,K_2)$-plane,  referred to as the $E$-plane in what follows. The left panel of Fig.~\ref{tri} shows the computed   $(K_1,K_2)$ values, plotted in the $E$-plane for an ordered grain of $225$ atoms ($87$ core atoms). Throughout this work, we use $J=0.25$~eV, a value consistent with a Kondo temperature below 0.1 K.  
 Different colors denote the MA parameters of clusters with magnetic impurities placed on the different shells. 
Similarly, the anisotropy constants   $K_{3}$, $K_{4}$, and $K_{5}$  (shown in Fig. 4 in  Ref. \cite{appendix}), induced by the SO interaction on the \textit{host} Au atoms, are related for 
atoms on the same shell, and show regular patterns  in a three dimensional space, the $T_{2}$-space. 
 These parameters are, however,  smaller by about one order of magnitude compared to the parameters 
 $K_{1,2}$,  implying that the MA constant, Eq. (\ref{K-LSO2}), is dominated by the $E$-type parameters.

Next, let us  examine the distribution of the magnetic anisotropy constants in case of disordered nanoclusters. First we created 100 disordered nanoclusters by 
adding 11 extra atoms to an 225-atom ordered cluster,  and placing them  randomly on the next shell of  24 possible sites. 
 We then calculated the MA parameters on all   core sites for every nanoparticle.  In Fig.~\ref{tri} (right) we show the 8700
 $E$-plane parameters obtained this way.  Different colors represent data from different shells.  Small 'clouds' are observed 
 with obvious remains of the three-fold symmetry, but there is no  strictly ordered structure left anymore in the $E$-plane.

\begin{figure}[t]
\begin{center}
\begin{tabular}
[c]{c}%
\includegraphics[width=0.5\textwidth, bb= 5 0 559 370 ]{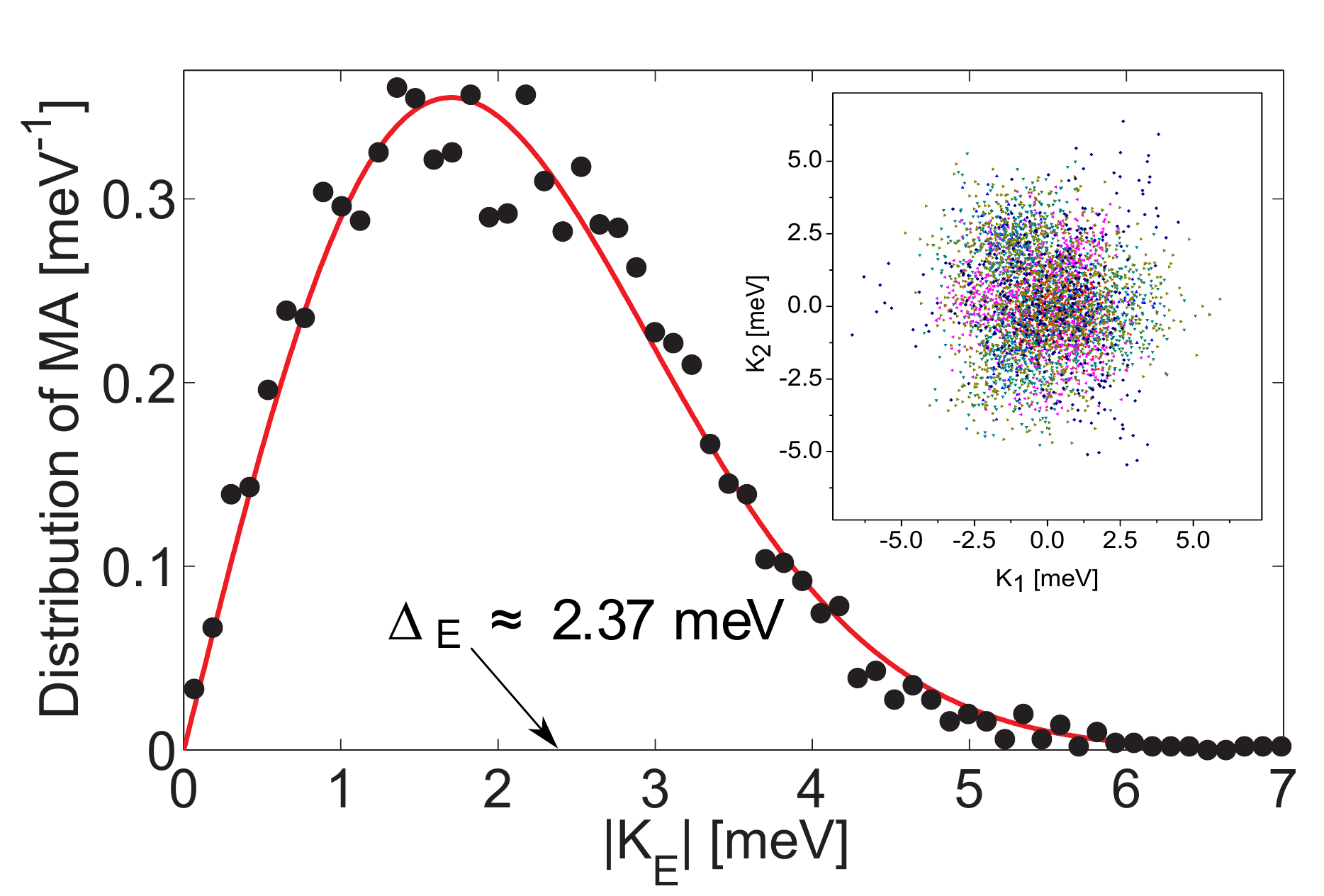}
\end{tabular}
\end{center}
\par
\vskip -0.5cm  
\caption{(color online) Radial distribution of the magnetic anisotropy parameters (dots) in $E$-plane in case of $N_{S}$=50 samples with N=225+25 atoms. The continuous line presents the predictions of the  Gaussian Orthogonal Ensemble. 
Inset: Distribution of $(K_1,K_2)$ in the $E$-plane for these 50 nano-grains. At this level of disorder the triangular structure is almost entirely lost.}
\label{extra25}
\end{figure}

We then increased the structural disorder of the nano-grains further, and added 25 extra atoms to an ordered cluster of 225 atoms,  by placing them
 randomly on the next three shells.  As shown in the inset  of Fig. \ref{extra25}, for these strongly disordered clusters  the distribution becomes almost  isotropic 
in the  $E$-plane, and the triangular symmetry is almost entirely  lost.  The main panel of Fig. \ref{extra25}  shows the radial distribution of the magnetic anisotropy parameters, $\sqrt{K_1^2 + K_2^2}$ in the $E$-plane.  The observed distribution agrees very well with the predictions of a simple Gaussian  theory, where we 
assume that the components of $\mathbf{K}_E  \equiv (K_1,K_2)$  have independent and Gaussian  distribution,
\begin{equation}
p(\mathbf{K}_E ) \sim e^{- \mathbf{K}_E^2/\Delta_E^2 } \;,
\label{eq:p(K)}
\end{equation}
with the E-plane anisotropy scale $\Delta_E $ defined as $\Delta^2_E \equiv \langle \mathbf{K}_E^2 \rangle$.
We find that  in these disordered grains the radial distribution in the three-dimensional $T_{2}$-space parameters can also be  fitted by a similar 
 Gaussian ensemble, although with a smaller  characteristic radius, $\sqrt{\langle \mathbf{K}_T^2 \rangle}\equiv \Delta_T < \Delta_E$ 
 (see Ref.~\cite{appendix}). 
The overall distribution of the anisotropy $K_L$ is therefore strongly suppressed at small values. In the absence of host SO coupling, 
 it scales as $p(K_L)\sim K_L$ for small anisotropy values, $K_L < \Delta_E$, while 
  in the presence of it $p(K_L)$ is suppressed as $p(K_L)\sim |K_L|^4 $ for $K_L<\Delta_T$. 
This implies that \emph{typical} sites in a disordered grain  have a \emph{finite} SO-induced  anisotropy of size  $\sim\Delta_E$, 
and of random orientation, almost independently of their precise location within the grain.  
 
We  remark that, -- even after adding a single extra atom to an ordered cluster, -- the distribution of 
 the eigenenergies of the host Hamiltonian  agreed
with the predictions of random matrix theory and, in agreement with the experimental findings~\cite{levelexp},  
exhibited level repulsion according to a Gaussian symplectic (GS) level spacing distribution. 
The observed GS distribution reflects the chaotic nature of the electron's wave function
as well as  the presence of  \emph{host} SO coupling \cite{appendix}. Building upon the chaotic nature of the electron's wave function, one can obtain an estimate of $\Delta_{E}$, using Eq.~\eqref{rho-int} and assuming random plane wave  
conduction electron wave functions \cite{berry}. This yields the estimate,
\begin{equation}
\Delta_E\sim J \;({\Delta_{SO}}/{\epsilon_F}) / {N^{3/2}},
\nonumber
\end{equation}
 with  $N$ the number of lattice sites on the cluster and
 $\Delta _{SO} $  the SO splitting of the $j=3/2$ and $j=5/2$   impurity-levels. 

\begin{figure}[b]
\begin{center}
\begin{tabular}
[c]{c}%
\includegraphics[width=0.5\textwidth, bb= 0 -10 630 370]{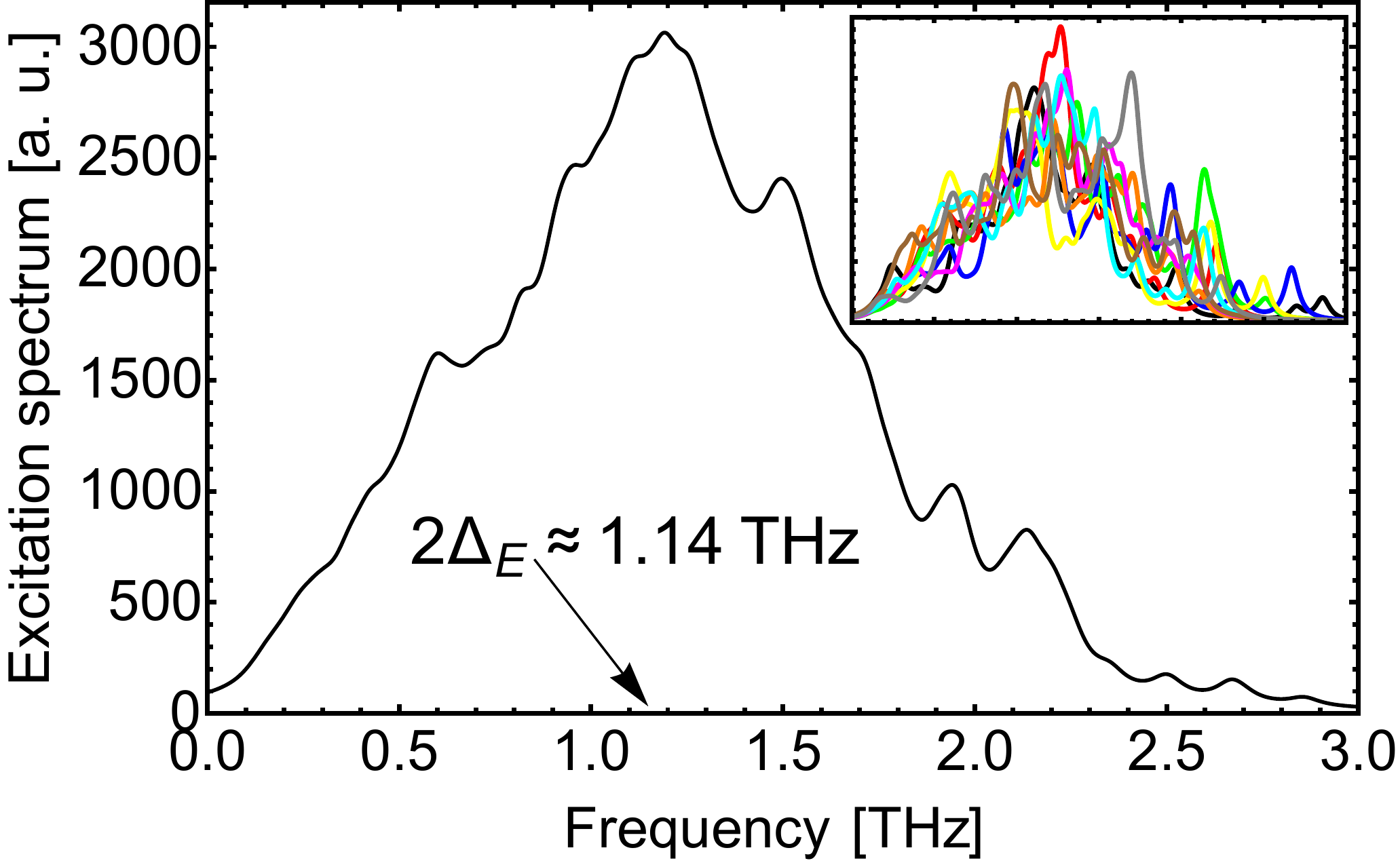}
\end{tabular}
\end{center}
\par
\vskip -0.5cm  \caption{(color online)
Orientational and disorder averaged excitation  spectrum computed 
for an ensemble of   1000 randomly chosen and randomly oriented disordered  nano-grains (N=225+40).
The shape of the signal reflects the distribution of magnetic  anisotropies while the small peak corresponds to transitions between the lowest Kramers doublets, split by the magnetic field.  Inset:  spectra of 100-100 selected grains \cite{appendix}.}%
\label{ESR}%
\end{figure}

The anisotropy distribution \eqref{eq:p(K)} has a direct impact on the magnetic excitation spectrum of the nano-grains. Generating 50 (strongly) disordered nano-grains with 225+40 atoms (87 core sites), we randomly chose and rotated 100 nano-clusters from the 4350 different samples (see Ref. \cite{appendix}), and determined the magnetic impurities' excitation spectrum averaging over the orientation of them. All these 100 spectra were added together, and the procedure was repeated ten times. 
The  obtained aggregated 
spectrum is shown in Fig.~\ref{ESR}. The obtained spectra are grain specific (see inset), and reflect directly the MA energy's distribution. The typical anisotropy values, $\Delta_{E} \approx 0.57$ THz, shift rapidly towards smaller values (GHz) with increasing grain size.

The universal anisotropy distribution should be indirectly observable through thermodynamic quantities, too. In the presence of a random distribution
of anisotropies, given by Eq.~\eqref{eq:p(K)}, we obtain a peak in the specific heat $C(T)$ at $T\approx 0.78 \; \Delta_E$, and a low temperature specific heat   $C(T)\sim T^2/\Delta_E^2$ (see Ref. \cite{appendix}), turning into a $\sim T^5$ anomaly for $T\ll \Delta_T$. 
Similarly, the coefficient of the Curie susceptibility, $\sim T\chi$,  should exhibit a strong suppression below $T\sim \Delta_E$.

Although the model discussed here has features that are specific, we believe that it captures many generic properties of
 magnetic impurities in a metallic grain, and thus allows one to draw general conclusions; For any magnetic impurity of spin $J$,  time reversal 
 symmetry implies that the leading anisotropy term is of the form \eqref{sop}. In ordered grains, the distribution of the  five parameters 
 $K_\mu$ must always  reflect the underlying lattice 
 symmetry, and for cubic lattices, in particular, the couplings ${\bf K}_E$ and ${\bf K}_T$  are organized
  into triangular and cubic structures, respectively. These parameters are expected to become random, and to exhibit 
multidimensional Gaussian distributions in sufficiently disordered grains. The suppression of the probability of having a 
small anisotropy, $p(K_L\to 0)= 0$, as well as the predicted specific heat and susceptibility anomalies are also  generic features, since they 
follow simply from the presence of  randomly distributed independent anisotropy parameters, $K_\mu$. Our conclusions regarding the 
Schottky anomaly are thus general, though details of the low temperature scaling of $C(T)$ may be system specific.

We owe thanks to Agnes Antal, Imre Varga and Jan Rusz for the fruitful discussions. This work has been financed by the Swedish Research Council, the KAW foundation, and the ERC (project 247062 - ASD) and the Hungarian OTKA projects K105148 and K84078. 
We also acknowledge support from eSSENCE and the Swedish National Allocations Committee (SNIC/SNAC).

\end{document}


\title{Universal distribution of magnetic anisotropy of impurities in ordered and disordered nano-grains}
\author{A. Szilva}
\affiliation{Department of Physics and Astronomy, Division of Materials Theory, Uppsala University,
Box 516, SE-75120, Uppsala, Sweden}
\affiliation{Department of Theoretical Physics, Budapest University of Technology and Economics, Budafoki \'ut 8. H-1111 Budapest, Hungary}
\author{P. Balla}
\affiliation{Department of Theoretical Physics, Budapest University of Technology and Economics, Budafoki \'ut 8. H-1111 Budapest, Hungary}
\affiliation{Institute for Solid State Physics and Optics, Wigner Research Centre for Physics,
Hungarian Academy of Sciences, H-1525 Budapest, P.O.B. 49, Hungary}
\author{O. Eriksson}
\affiliation{Department of Physics and Astronomy, Division of Materials Theory, Uppsala University,
Box 516, SE-75120, Uppsala, Sweden}
\author{G. Zar\'and}
\affiliation{BME-MTA Exotic Quantum Phases 'Lend\"ulet' Group, Institute of Physics, Budapest University of Technology and Economics, H-1521 Budapest, Hungary}
\author{L. Szunyogh}
\affiliation{Department of Theoretical Physics, Budapest University of Technology and Economics, Budafoki \'ut 8. H-1111 Budapest, Hungary}
\affiliation{MTA-BME Condensed Matter Research Group, Budapest University of
Technology and Economics, Budafoki ´ut 8., H-1111 Budapest, Hungary}

\begin{abstract}

\end{abstract}
\pacs{later}

\maketitle

\section{Appendix A - Description of the gold nanograin-host}

We have defined the structure of the Au nano-grains host of $N$=100-400 atoms as follows: one can speak of an \textit{ordered} grain when it has only filled shells around a \textit{center} atom, while nano-particles with partially filled (outmost) shells have been referred to as \textit{disordered} nano-grains. The \textit{shell} is a group of atoms ($N_{sh}$) on an \textit{fcc} lattice that transform into each other under the cubic group ($O_{h}$). The shell-structure of a few fcc nano-clusters is shown in Table \ref{tablegrain}: an ordered grain built by, say, $N$=225 atoms has $12$ filled shells: a $1$ center atom, $12$ first-, $6$
second- ... and $24$ twelfth-neighbors. The site $(0,1.5,1.5)$ belongs to shell $9NNa$, while the site $(0.5,0.5,2)$ is on shell $9NNb$, though they are at the same distance from the origin (center atom).  If a disordered grain host of $N=236$ atoms has $11$ atoms in the outmost ($12NN$) shell (instead of $N_{sh}=$24 atoms), then one can deal with $C^{24}_{11}$ configurations. In practice, we choose randomly only $N_{S}$ grains from the big configuration space ($N_{S}$ is typically set as 50-100). It should be noted that one can put extra atoms not only into the first outmost shell but we never generate "holes" in a nano-cluster. In a given nano-grain the atoms those that have all the first neighbors are called as $core$ atoms (denoted by $N_{c}$ in Table \ref{tablegrain}). The core region is away from the surface of the nano-grain.  If a nano-particle is built by $N$=225 atoms then it has $N_{c}$=87 cora atoms (being in the center site and on six core shells). 

\begin{table}[h]
\caption{The shell structure of fcc clusters. Label, $a$ and $b$
denote the shells where the atoms are at the same distance from the center atom but cannot be transformed into each other (under $O_{h}$). $N_{sh}$ denotes
the number of sites in a given shell, N is the total number of atoms and $N_{c}$ is the number of core sites in the cluster.}%
\label{tablegrain}
\begin{center}
\begin{tabular}
[c]{|lllllllll|}\hline
&  &  &  &  &  &  &  & \\
Shell & $N_{sh}$ & $N$ & $N_{c}$ &  & Shell & $N_{sh}$ & $N$ & $N_{c}$\\
&  &  &  &  &  &  &  & \\
$Center$ & \;$1$ & \;$1$ & \;$0$ & \; & \;$16NNa$ & \;$24$ & \;$405$ &
\;$177$\\
$1NN$ & \;$12$ & \;$13$ & \;$1$ & \; & \;$16NNb$ & \;$24$ & \;$429$ &
\;$201$\\
$2NN$ & \;$6$ & \;$19$ & \;$1$ & \; & \;$17NNa$ & \;$24$ & \;$453$ & \;$225$\\
$3NN$ & \;$24$ & \;$43$ & \;$1$ & \; & \;$17NNb$ & \;$6$ & \;$459$ & \;$225$\\
$4NN$ & \;$12$ & \;$55$ & \;$13$ & \; & \;$18NN$ & \;$48$ & \;$507$ &
\;$249$\\
$5NN$ & \;$24$ & \;$79$ & \;$19$ & \; & \;$19NN$ & \;$24$ & \;$531$ &
\;$249$\\
$6NN$ & \;$8$ & \;$87$ & \;$19$ & \; & \;$20NN$ & \;$24$ & \;$555$ & \;$273$\\
$7NN$ & \;$48$ & \;$135$ & \;$43$ & \; & \;$21NN$ & \;$48$ & \;$603$ &
\;$321$\\
$8NN$ & \;$6$ & \;$141$ & \;$43$ & \; & \;$22NN$ & \;$24$ & \;$627$ &
\;$321$\\
$9NNa$ & \;$12$ & \;$153$ & \;$55$ & \; & \;$23NN$ & \;$48$ & \;$675$ &
\;$369$\\
$9NNb$ & \;$24$ & \;$177$ & \;$55$ & \; & \;$24NN$ & \;$8$ & \;$683$ &
\;$369$\\
$10NN$ & \;$24$ & \;$201$ & \;$79$ & \; & \;$25NNa$ & \;$48$ & \;$731$ &
\;$393$\\
$11NN$ & \;$24$ & \;$225$ & \;$87$ & \; & \;$25NNb$ & \;$12$ & \;$743$ &
\;$405$\\
$12NN$ & \;$24$ & \;$249$ & \;$87$ & \; & \;$26NN$ & \;$24$ & \;$767$ &
\;$411$\\
$13NNa$ & \;$48$ & \;$297$ & \;$135$ & \; & \;$27NN$ & \;$24$ & \;$791$ &
\;$435$\\
$13NNb$ & \;$24$ & \;$321$ & \;$141$ & \; & \;$28NN$ & \;$24$ & \;$815$ &
\;$459$\\
$14NN$ & \;$48$ & \;$369$ & \;$165$ & \; & \;$29NN$ & \;$24$ & \;$839$ &
\;$459$\\
$15NN$ & \;$12$ & \;$381$ & \;$177$ & \; & \; & \; & \; & \;\\
&  &  &  &  &  &  &  & \\\hline
\end{tabular}
\end{center}
\end{table}

The electronic structure of the Au nano-grains will be described by a tight binding (TB) Hamiltonian. The model uses $spd$ canonical orbitals, and the spin-orbit (SO) coupling of the host atoms is considered non-perturbatively. Specifically, the TB model uses (nearly)
orthonormal basis functions which are localized at sites, $\mathbf{R}_{n}$,
\begin{equation}
\langle\mathbf{r}\mid n;\alpha\sigma\rangle=\langle\mathbf{r}-\mathbf{R}_{n}%
\mid\alpha\sigma\rangle=\psi_{\alpha}(\mathbf{r}-\mathbf{R}_{n})\phi_{\sigma}\;,
\label{bazfv}%
\end{equation}
where $n$ refers to the given site,
the index $\alpha$ denotes the so-called canonical basis (real spherical harmonics),
\begin{equation}%
\begin{tabular}
[c]{ll}%
$\alpha=s$ & $\ell=0$\\
$\alpha=p_{x},p_{y},p_{z},$ & $\ell=1$\\
$\alpha=d_{xy},d_{xz},d_{yz},d_{x^{2}-y^{2}},d_{3z^{2}-1}$ & $\ell=2$%
\end{tabular}
\ \ \ \ \;, \label{kan}%
\end{equation}
$\psi_{\alpha}$ depends only on the azimuthal quantum number $\ell$ and the
spin quantum number is labeled by $\sigma=\pm\frac{1}{2}$.

The Hamiltonian of the noble metal host is written as
\begin{equation}
\hat{H}=\left\lbrace H_{\alpha\sigma,\alpha^{\prime}\sigma^{\prime}%
}^{n,n^{\prime}}\right\rbrace =(\varepsilon_{\alpha}\delta_{\alpha\alpha^{\prime}}%
\delta_{\sigma\sigma^{\prime}}+\xi H_{\alpha\sigma,\alpha^{\prime}%
\sigma^{\prime}}^{LS})\delta_{nn^{\prime}}+t_{\alpha,\alpha^{\prime}%
}^{n,n^{\prime}}\delta_{\sigma\sigma^{\prime}}\;, \label{hamclus}%
\end{equation}
where the dimension of the matrix is $M=18 \times N$, $\varepsilon_{\alpha}$ is the so-called on-site energy parameter,%
\begin{equation}
H_{\alpha\sigma,\alpha^{\prime}\sigma^{\prime}}^{LS}=\langle\,\alpha
\sigma|\,\vec{L}\,\vec{S}\,|\,\alpha^{\prime}\sigma^{\prime}\rangle\;,
\end{equation}
$\xi$ is the SO coupling parameter and $t_{\alpha,\alpha^{\prime}%
}^{n,n^{\prime}}$are the hybridization matrix elements (or hopping
integrals) between the different orbitals. 

We note that on-site energies $\varepsilon_{s}$, $\varepsilon_{p}$, $\varepsilon_{d-E_{g}}$  and $\varepsilon_{d-T_{2g}}$ were used in case of all calculations. The hopping integrals to first- and second
nearest neighbors were considered. They depend only on the relative positions of the
sites, i.e.,
\begin{equation}
t_{\alpha,\alpha^{\prime}}^{n,n^{\prime}}=t_{\alpha,\alpha^{\prime
}}\left(  \mathbf{R}_{n^{\prime}}-\mathbf{R}_{n}\right)  \;.
\label{pot}%
\end{equation}
The numerical values for both $\varepsilon_{\alpha}$ and $t_{\alpha,\alpha^{\prime
}}$ can be found in Ref. \cite{doktori}. The matrixelements of the SO coupling can easily be calculated with the help of
following identity,
\begin{equation}
\vec{L}\,\vec{S}=\frac{1}{2}\left(  L_{+}S_{-}+L_{-}S_{+}\right)  +L_{z}S_{z}\;.%
\end{equation}
The spin-orbit coupling parameter $\xi$ was determined from
the difference of the SO-split $d$-resonance energies%
\begin{equation}
\Delta E_{d}=E_{j=5/2}-E_{j=3/2}\;,
\end{equation}
as derived from self-consistent relativistic (SKKR) first-principles calculations
\cite{pB22}. This splitting is related to the strength of SO coupling as
\begin{equation}
\Delta E_{d}\simeq\frac{5}{2}\xi\;.
\end{equation}
For Au bulk we obtained $\xi=0.64$ $eV$. 

The Green's function or resolvent operator of a nano-particle is defined as
\begin{equation}
\hat{G}(z)=G_{\alpha\sigma,\alpha^{\prime}\sigma^{\prime}%
}^{n,n^{\prime}}(z)=(z-\hat{H})^{-1}\;\label{G1}%
\end{equation}
that can be written as
\begin{equation}
\hat{G}(z)=\sum_{i=1}^{M}\frac{|v_{i}\rangle\langle v_{i}%
|}{z-\varepsilon_{i}}\;, \label{G2}%
\end{equation}
where $\{\varepsilon_{i}\}$ and $\{|v_{i}\rangle\}$ stand for the eigenvalues and eigenvectros of the Hamiltonian Eq. (\ref{hamclus}), respectively. 

\begin{figure}[h!]
\begin{center}
\begin{tabular}
[c]{c}%
\includegraphics[width=0.8\textwidth, bb=  20 20 1200 570]{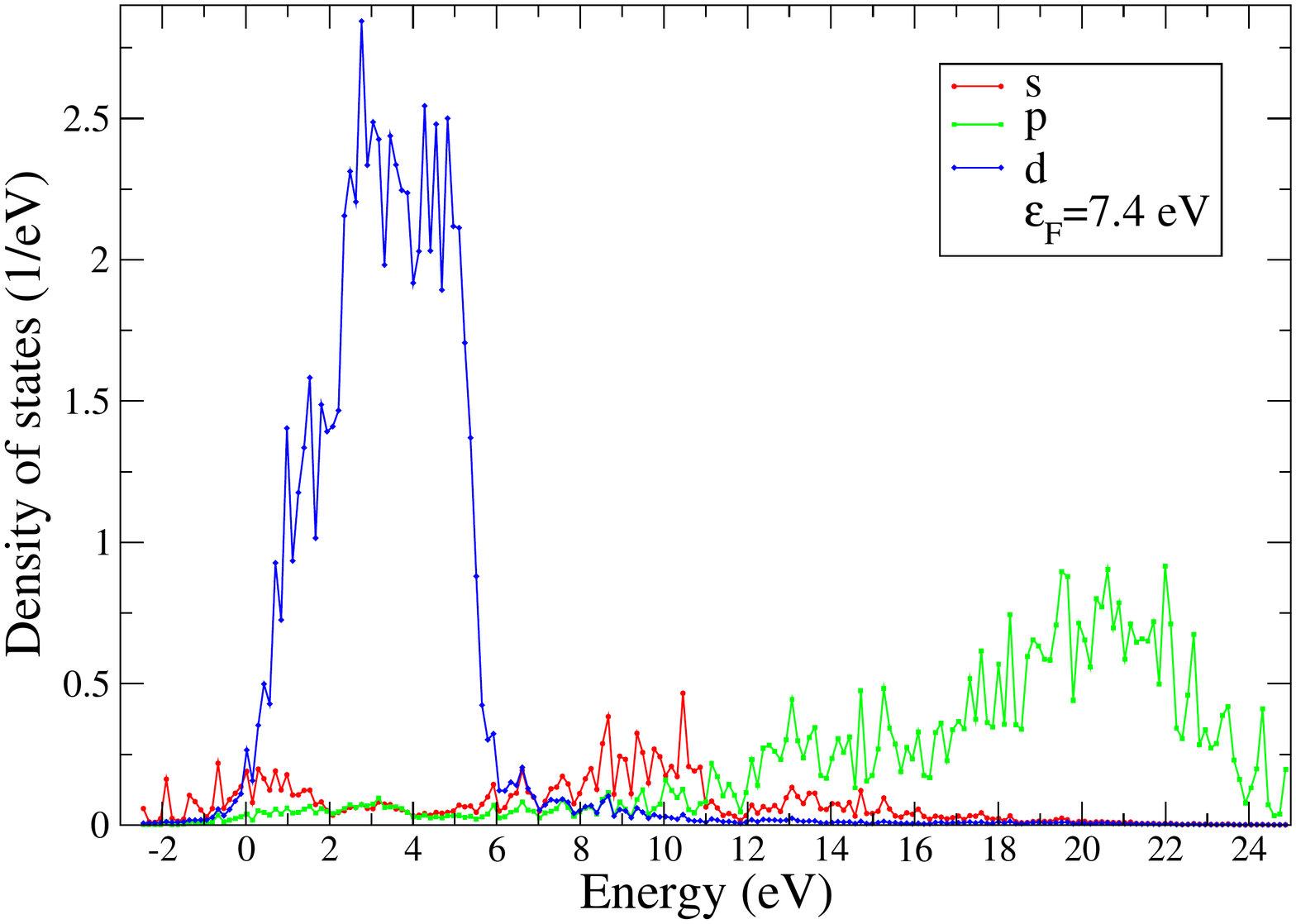}
\end{tabular}
\end{center}
\par
\vskip -0.5cm  \caption{(color online) The $spd$ density of states (DOS) components of an
ordered nano-grain (N=225) normalized to one atom. The calculated Fermi-energy is $\varepsilon_{F}=7.4$ eV.}%
\label{spd1}%
\end{figure}

Finally we define the density of states (DOS) as follows,
\begin{align}
n (\varepsilon)  &  =-\frac{1}{2\pi i}\lim_{\delta
\rightarrow0}Tr\left(  \hat{G}(\varepsilon+i\delta)-\hat{G}(%
\varepsilon-i\delta)\right)  \;. \label{hostDOS}
\end{align}
The numerically calculated values are shown in Fig. \ref{spd1} for an ordered nano-grain. The calculated Fermi-energy is $\varepsilon_{F}=7.4$ eV, at the Fermi-energy the $s$ contribution dominates the DOS.

\section{Appendix B - Derivation of the LSO model Hamiltonian}

Next we derive the local spin-orbit (LSO) model for a $d^{1}$-type magnetic impurity.  In the (non-degenerate) Anderson impurity model \cite{anderson} a single energy level, $\varepsilon_{d}$, is considered at the impurity. This level can be at most doubly occupied. Because of the Coulomb repulsion, the energy of the doubly occupied state is $2\varepsilon_{d}+U$. Moreover, since the wavefunction of the $d$ level is not orthogonal to the states of the conduction band, they may be hybridized. We consider a free ion that has the following four possible states: the $d$ level is empty
in state $|d^{0}\rangle$, it is occupied by an electron with spin $\sigma$ 
in state $|d^{1}_{\sigma}\rangle$, and it is doubly occupied by electrons with opposite spins 
in state $|d^{2}\rangle$. The energies of these states are $\varepsilon(d^{0})=0$,
 $\varepsilon(d^{1}_{\sigma})=\varepsilon_{d}$ and $\varepsilon(d^{2})=2\varepsilon_{d}+U$.
The ground state of the ion has a magnetic moment if the magnetic doublet
is the lowest in energy, that is if $\varepsilon(d^{2})>\varepsilon(d^{1}_{\sigma})$
and $\varepsilon(d^{0})>\varepsilon(d^{1}_{\sigma})$, i.e. if
$-\frac{1}{2}U<\varepsilon_{d}+\frac{1}{2}U<\frac{1}{2}U$.
When $U$ is large, no double occupancy happens, and the spin fluctuates.

The so-called $ionic$ $model$ is a possible generalization of the
Anderson model \cite{hirst}. The
Hamiltonian contains three terms as usual: the \textit{impurity},
the \textit{host} (conduction) electron, and the
\textit{hybridization} terms. If the multiplet of 
the impurity ion is denoted by $|n,m_{n}\rangle$, where $n$
indicates the number of electrons in the shell and $\{m_{n}\}$ the set of
quantum numbers characterizing the multiplet, and the corresponding energy
is denoted by $E_{n,m_{n}}$, then the impurity term can be written as
\begin{equation}
\mathcal{H}_{imp}=\sum_{n,m_{n}}E_{n,m_{n}}|n,m_{n}\rangle\langle n,m_{n}|\;.
\end{equation}
If the impurity has only a single non-degenerate $d$ level and 
the Coulomb interaction $U$ is sufficiently  large, so that the doubly
occupied configuration has high energy, so that the only relevant configurations are $n=0$ or $n=1$, then the Hamiltonian of the ionic model, $\mathcal{H}_{ionic}$, can be approximated as
\begin{align}
E_{0}|0,0\rangle\langle0,0|+ \sum_{m}E_{1,m}|1,m\rangle
\langle1,m|+  \sum_{\vec{k},m}\varepsilon_{\vec{k}}s_
{\vec{k},m}^{\dagger}s_{\vec{k},m}+\nonumber\\
 +\sum_{\vec{k},m}\left(  V_{\vec{k}}|1,m\rangle\langle0,0|
s_{\vec{k},m}+V_{\vec{k}}^{\ast}%
s_{\vec{k},m}^{\dagger}|0,0\rangle\langle1,m|\right)  \;, \label{ionic}%
\end{align}
where $m$ runs over $2j+1$ values. In case of $j=\frac{1}{2}$ the
$U=\infty$ Anderson model is recovered, see Ref. \cite{hewson}, Sec. 1.9. In Eq. (\ref{ionic}) the operator $s_{\vec{k},m}^{\dagger}$ creates a host conduction electron with
wavenumber $\vec{k}$, pseudospin $m$ and energy $\varepsilon_{\vec{k}}$. $V_{\vec{k}}$-s denote the $s$-$d$ hybridization matrix elements.

As we mentioned in the main paper, the desired effective Hamiltonian, which describes the interaction of the magnetic impurity and the host (conduction) electrons, should
be invariant under the cubic group symmetry. Therefore we have to use the $\Gamma_{8}$ symmetry adapted combinations of the host $s$ orbitals ($d$-type or, more precisely, $E$-type combination of these orbitals).

The gold host atoms form an $fcc$ lattice: an
impurity has $12$ nearest neighbor host atoms (in the core region of a nano-grain). The operators $s_{xy}^{\dagger}$, $s_{\bar{x}y}^{\dagger}$, $s_{x\bar{y}}^{\dagger}$, $s_{\bar{x}\bar{y}}^{\dagger}$, $s_{xz}^{\dagger}$, $s_{yz}^{\dagger}$, $s_{\bar{x}z}^{\dagger}$, $s_{\bar{y}z}^{\dagger}$,$s_{x\bar{z}}^{\dagger}$, $s_{y\bar{z}}^{\dagger}$, $s_{\bar{x}\bar{z}}^{\dagger}$, $s_{\bar{y}\bar{z}}^{\dagger}$ create the appropriate $s$-electrons at the nearest neighbor sites with either spin $\sigma$. The subscript of the creation operators refer to the nearest neighbor sites. For instance if the impurity takes place at site ($0,0,0$) then index $xy$ denotes  the coordinate of the site ($\frac{1}{2},\frac{1}{2},0$). Using the combinations
\begin{align}
s_{xy}^{e}  &  =\frac{1}{\sqrt{2}}\left(  s_{xy}^{\dagger}+s_{\bar{x}\bar{y}%
}^{\dagger}\right)  \;,\;s_{\bar{x}y}^{e}=\frac{1}{\sqrt{2}}\left(  s_{\bar
{x}y}^{\dagger}+s_{x\bar{y}}^{\dagger}\right)  \;,\nonumber\\
s_{xz}^{e}  &  =\frac{1}{\sqrt{2}}\left(  s_{xz}^{\dagger}+s_{\bar{x}\bar{z}%
}^{\dagger}\right)  \;,\;s_{\bar{x}z}^{e}=\frac{1}{\sqrt{2}}\left(  s_{\bar
{x}z}^{\dagger}+s_{x\bar{z}}^{\dagger}\right)  \;,\nonumber\\
s_{yz}^{e}  &  =\frac{1}{\sqrt{2}}\left(  s_{yz}^{\dagger}+s_{\bar{y}\bar{z}%
}^{\dagger}\right)  \;,\;s_{\bar{y}z}^{e}=\frac{1}{\sqrt{2}}\left(  s_{\bar
{y}z}^{\dagger}+s_{y\bar{z}}^{\dagger}\right)  \;, \label{evens}%
\end{align}
the $d$-like combinations, $D_{_{\delta}}$, can be expressed as
\begin{align}
D_{xz}  &  =\frac{1}{\sqrt{2}}\left(  s_{xz}^{e}-s_{\bar{x}z}^{e}\right)
\;,\;D_{yz}=\frac{1}{\sqrt{2}}\left(  s_{yz}^{e}-s_{\bar{y}z}^{e}\right)
\; , \label{dorbit}%
\end{align}
etc.
Including spin variables this leads to ten $D_{_{\delta}}|\sigma\rangle$ combinations.
The $T_{2}$-type combinations $D_{xz}$, $D_{yz}$, $D_{xy}$ transform as $\Gamma_{5}$, while the $E$-type combinations $D_{x^{2}-y^{2}}$
and $D_{2z^{2-}x^{2}-y^{2}}$
transform as $\Gamma_{3}$. As we mentioned in the main paper, the $\Gamma_{3}$ combinations are relevant for the construction of the LSO effective Hamiltonian. To construct the LSO symmetry adapted quantities, we should find the relation between standard $\Gamma_{8}$ basis
\begin{equation}
\left(
\begin{array}
[c]{c}%
|s_{3/2}\rangle\\
|s_{1/2}\rangle\\
|s_{-1/2}\rangle\\
|s_{-3/2}\rangle
\end{array}
\right)  \;, \label{sop}%
\end{equation}
and the tensor-product basis $\left\{  D_{_{\delta}}|\sigma\rangle
\right\}$, where $\delta=x^{2}-y^{2}$ or $2z^{2}-x^{2}-y^{2}$ and $\sigma=\uparrow
,\downarrow$. We should find the transformation matrix between the two basis as follows,
\begin{equation}
\left\{  |s_{m}\rangle\right\}  =Q\left\{  D_{_{\delta}}|\sigma\rangle
\right\}  \;. \label{Qtrafo}%
\end{equation}
To find the proper $Q$, we compared the action of a rotation around the $z$ and $x$ axis by
angle $\frac{\pi}{2}$ in both bases and obtained 
\begin{equation}
Q=\left(
\begin{array}
[c]{cccc}%
0 & -1 & 0 & 0\\
0 & 0 & 1 & 0\\
0 & 0 & 0 & -1\\
1 & 0 & 0 & 0
\end{array}
\right)  \;, \label{Q}%
\end{equation}
implying
\begin{align}
|s_{3/2}\rangle &  =-D_{x^{2}-y^{2}}|\downarrow\rangle\;,\nonumber\\
|s_{1/2}\rangle &  =D_{2z^{2}-x^{2}-y^{2}}|\uparrow\rangle\;,\nonumber\\
|s_{-1/2}\rangle &  =-D_{2z^{2}-x^{2}-y^{2}}|\downarrow\rangle\;,\nonumber\\
|s_{-3/2}\rangle &  =D_{x^{2}-y^{2}}|\uparrow\rangle\;.
\label{ssorbit}%
\end{align}
By probing for the rest of the generating point group elements of
the cubic point-group it can easily be shown that this basis forms indeed
a $\Gamma_8$ representation of the cubic double point-group.

The ionic Hamiltonian, Eq. (\ref{ionic}), without kinetic energy of the host electrons reduces to
\begin{equation}
\mathcal{H}_{LSO}=E_{d}\sum_{m}|m\rangle\langle m|+V\sum_{m}\left(  |m\rangle\langle
0|s_{m}+s_{m}^{\dagger}|0\rangle\langle m|\right)  \;, \label{hprime}%
\end{equation}
where $V$ is the hybridization parameter, and we choose $E_{0}=0$ and
$E_{1,m}=E_{d}$ for each $m$. By using a Coqblin-Schrieffer
canonical transformation, see Refs. \cite{pB17} and \cite{hewson}, Sec. 1.10,
for the $\mathcal{H}_{LSO}$ Hamiltonian, Eq. (\ref{hprime}), we obtain that
\begin{equation}
\mathcal{H}_{LSO}=J\sum_{m,m^{\prime}}s_{m}^{\dagger}s_{m^{\prime}}X_{m^{\prime}m}\;, \label{hlso}%
\end{equation}
where the Hubbard operators, $X_{m^{\prime}m}=|m^{\prime}\rangle\langle m|$, refer to the states $\{\frac{3}{2},\frac
{1}{2},-\frac{1}{2},-\frac{3}{2}\}$ of the impurity, and $s_{m}^{\dagger}$ creates
appropriate host electrons, while $J$ denotes the exchange constant,
\begin{equation}
J=\frac{V^{2}}{|E_{d}|}\;, \label{jlso}%
\end{equation}
where $J$ was set as 0.25 eV (being consistent with the Kondo temperature below 0.1 K). This procedure is a natural generalization of the derivation of the $s$-$d$ model by Schrieffer and Wolff for case of $j=\frac{1}{2}$.

We note that the LSO Hamiltonian Eq. (\ref{hlso}) is basically the $U=\infty$ limit of the ionic model. Considering the large Coulomb interaction of the impurity $d$-level, the $n=2$ double occupation is not allowed and the above Hamiltonian cannot be obtained in this simple form. However, we are convinced that a finite U can result in a magnetic anisotropy (MA) of the same order of magnitude, i.e. the LSO model is the basic mechanism of the partially (but not half) filled magnetic impurities embedded into noble metal nanosystems. In Refs. \cite{kettes} and \cite{harmas}  Au(Fe) and Cu(Fe) reduced dimensional dilute magnetic alloy systems (thin films) were analyzed, while Ref. \cite{negyes} dealt with Cu(Mn) system. In case of Fe impurities, the suppression of the Kondo effect has been observed but this effect for Mn has not been measured yet. This observation is in agreement with our expectation that the LSO model can produce large enough magnetic anisotropy to explain the reduction of the Kondo effect, and suggests that the finite U does not have importance.

\section{Appendix C - Calculation of the magnetic anisotropy}

Here we derive the MA matrix from the LSO model. We should note that the host Hamiltonian, Eq. (\ref{hamclus}), must be modified in the presence of a
magnetic impurity that breaks the two-dimensional translation symmetry. The simplest way to account for this
constraint is to shift the on-site $d$-state energies of the impurity
$\varepsilon_{\alpha}^{i}$ far below the valence band and add the following
term to the Hamiltonian,
\begin{equation}
\Delta \hat{H}=\Delta H_{\alpha\sigma,\alpha^{\prime}\sigma
^{\prime}}^{n,n^{\prime}}=\left(  \varepsilon_{\alpha}^{i}-\varepsilon
_{\alpha}\right)  \delta_{n0}\delta_{n^{\prime}0}\delta_{\alpha\alpha^{\prime
}}\delta_{\sigma\sigma^{\prime}}\;, \label{dhgrain}%
\end{equation}
where the impurity is at site $n=0$ of a nano-grain. 

According to the LSO model, we need the Green's function only for a cluster of sites,
$\mathcal{C}$, consisting of nearest neighbor atoms around the impurity and of the impurity itself (12+1 atoms). The corresponding Green's function matrix can be evaluated as
\begin{equation}
\hat{g}(z)=\hat{g}^{\prime} (z) \left(  \hat{I}-\Delta \hat{H}^{\prime}\hat{g}(z)\right)  ^{-1}\;, \label{localgreen}%
\end{equation}
where $\hat{I}$ is a unit matrix and 
$\hat{g}^{\prime}(z)=\{\hat{G}(z)\}_{\mathcal{C}}$, and $\hat{G}(z)$ is defined by Eq. (\ref{G2}). The spectral function matrix of cluster $\mathcal{C}$ is then defined as%
\begin{align}
{\hat{\rho}}_{\mathcal{C}}(\varepsilon) &  =-\frac{1}{2\pi i}\lim_{\delta
\rightarrow0}\left(  \hat{g}(\varepsilon+i\delta)-\hat{g}
(\varepsilon-i\delta)\right)  \;. \label{rhohost}
\end{align}
The dimension of the matrices defined in Eqs. (\ref{localgreen}) and (\ref{rhohost}) is
 $(13\times9\times2)$, because the number of sites in the small cluster $\mathcal{C}$ is 13, and the orbital and spin-indices are $9$ and $2$, respectively. Our impurity
model is restricted to the hybridization between the impurity and the $s$-type
conduction electrons, hence, from the $s$-components of the
${\hat{\rho}}_{\mathcal{C}}(\varepsilon)$ matrix we define the following projected matrix,
\begin{equation}
\mathbf{\rho}^{nn'}_{s-\mathcal{C},\sigma \sigma'}
(\varepsilon)=\mathbf{\rho}^{nn'}_{\mathcal{C},s\sigma,s\sigma'}%
(\varepsilon)\;, \label{rhonn}%
\end{equation}
where $n,n' \in \mathcal{C}$ and $\hat{\rho}_{s-\mathcal{C}}$ is a $(2\times 13) \times (2 \times 13)$ matrix, considering the the up ($\uparrow$) and down ($\downarrow$) spin channels.

Then we construct a $4\times 4$ $\rho^{*}(\varepsilon)$ matrix from $\hat{\rho}_{s-\mathcal{C}}$ referring to its elements by $\delta_{1}$, $\delta_{2}$ and $\sigma$, $\sigma'$ indices, where $\delta_{1}=x^{2}-y^{2}$ and $\delta_{2}=2z^{2}-x^{2}-y^{2}$ are the $E$-type orbitals. Next we transform this matrix into the $\Gamma_{8}$ symmetry adapted basis by using Eq. (\ref{Q}) as follows,
\begin{equation}
\rho^{L}(\varepsilon)=Q\rho^{*}(\varepsilon)Q^{\dagger}\;.
\label{rtrafo}
\end{equation}

\begin{figure}[h!]
\begin{center}
\begin{tabular}
[c]{c}%
\includegraphics[
width=8cm,bb=0 0 415 165]{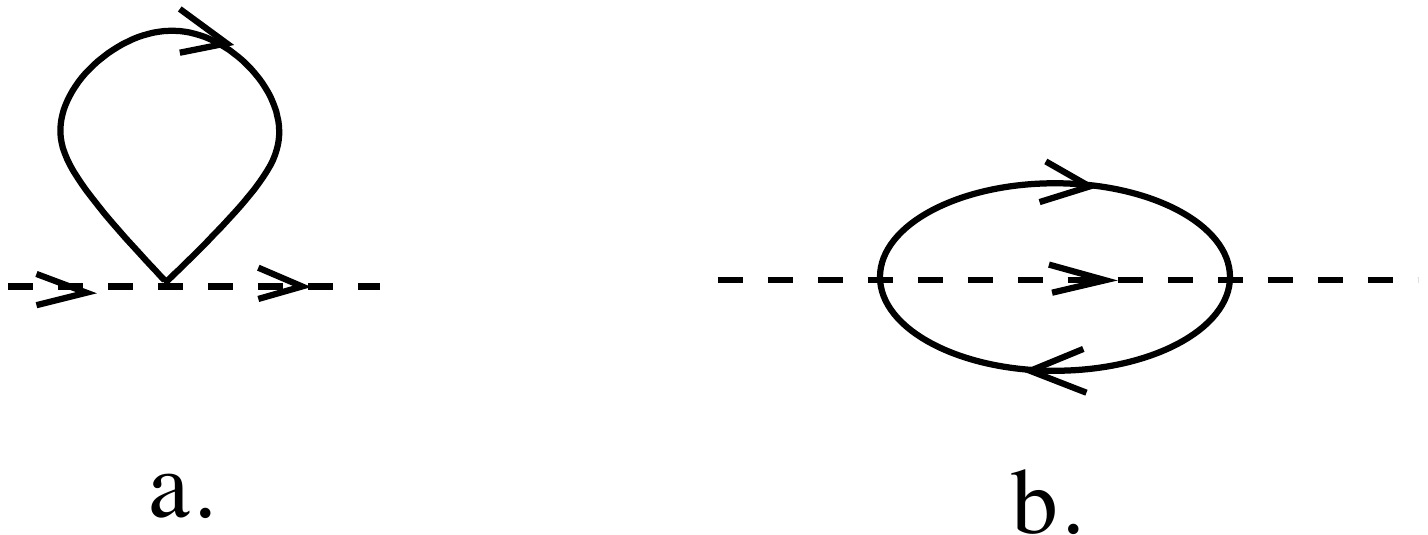}
\end{tabular}
\end{center}
\par
\vskip -0.5cm \caption{First and second ordered self
energy diagrams of the impurity spin. The dashed and continuous lines denote the
propagators of the spin and the conduction electrons, respectively. }%
\label{graf}
\end{figure}

To calculate the splitting of the four states, we employ Abrikosov's pseudofermion representation \cite{abri}. Using the up to second order in $J$, the diagrams are shown in Fig. \ref{graf}, and the self-energy at $T=0$ temperature is given by
\begin{equation}
\Sigma_{m\;m^{\prime}}(\omega=0)=\Sigma_{m\;m^{\prime}}^{(1)}%
+\Sigma_{m\;m^{\prime}}^{(2)}\;,
\end{equation}
where
\begin{equation}
\Sigma_{mm^{\prime}}^{(1)}=J\int_{-\infty}^{\varepsilon_{F}}d\varepsilon
\rho^{L}_{mm^{\prime}}(\varepsilon)\;, \label{rho-int}%
\end{equation}%
and
\begin{equation}
\Sigma_{mm^{\prime}}^{(2)}=J^{2}\int_{-\infty}^{\varepsilon_{F}}%
d\varepsilon\int_{\varepsilon_{F}}^{\infty}d\varepsilon^{\prime}\frac
{1}{\varepsilon^{\prime}-\varepsilon}\rho^{L}_{mm^{\prime}}(\varepsilon)%
{\displaystyle\sum\limits_{m^{\prime\prime}}}
\rho^{L}_{m^{\prime\prime}m^{\prime\prime}}(\varepsilon^{\prime})\;,
\end{equation}
where $\rho^{L}_{mm^{\prime}}(\varepsilon)$ are the elements of $\rho^{L}(\varepsilon)$ computed in the absence of the exchange interaction, i.e. $J=0$, and $\varepsilon_{F}$ is the Fermi-energy \cite{pB12}. Interestingly, already the first-order contribution to the self-energy gives a
nonvanishing anisotropy in the vicinity of a surface or at a site of a
nano-grain. Therefore, as what follows we consider this term only. In the main paper we identify the MA matrix as the resulted first order self-energy,
\begin{equation}
\left\lbrace K_{mm^{\prime}}\right\rbrace=\left\lbrace \Sigma_{mm^{\prime}}^{(1)}\right\rbrace \;,
\end{equation}
and the effective spin Hamiltonian can be written as
\begin{equation}
H^{L}= \sum_{m,m'} K_{mm^\prime}|m^{\prime} \rangle\langle m| \;.
\label{HL}
\end{equation}

\begin{figure}[h!]
\begin{center}
\begin{tabular}
[c]{c}%
\includegraphics[width=0.8\textwidth, bb=25 125 930 560]{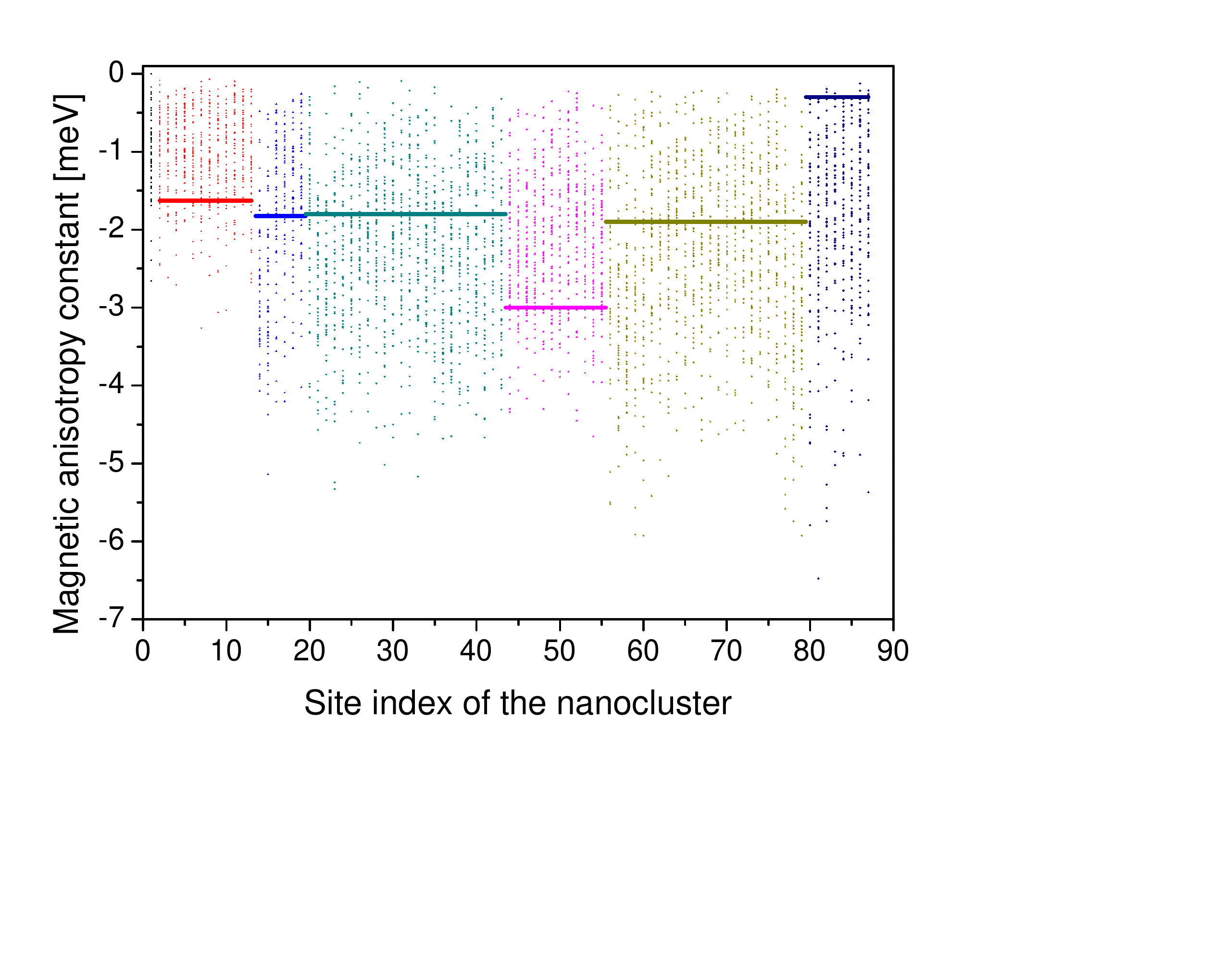}
\end{tabular}
\end{center}
\par
\vskip -0.5cm  \caption{(color online) Calculated MA constant, $K_{L}$, values for both ordered and disordered nano-balls. The ordered cluster has $N$=225 atoms, the solid lines correspond to the values of $K_{L}$ calculated for the ordered case. The number of core sites is $N_{c}$=87, i.e. the MA is for atoms located in the center site and on the first six core shells, see the different colors. The number of all sites in disordered nano-balls is $N$=265, i.e. 40 extra atoms are put to the next three outmost shells. The number of sample is $N_{S}=50$.}%
\label{MAfig}%
\end{figure}

The structure of $H^{L}$ is given by Eq. (3) in the main paper, the MA matrix can be parametrized by five real numbers, $K_{\mu}$ ($\mu=1...5$). The difference of its Kramers-degenerate eigenvalues defines the LSO magnetic anisotropy constant $K_{L}$ (see Eq. (4) in the main paper). The MA constant should be the same for sites in the same shell because of the $O_{h}$ symmetry relations. The thick solid lines in Fig. \ref{MAfig} show the corresponding MA values for different (core) shells in case of an ordered nano-grain with 225 atoms. If one randomly puts 40 extra atoms on the sites of the next three outmost shells, constructing disordered nano-grains, then the MA is not shell-degenerate anymore.

It should be also noted that in case of tetragonal symmetry, e.g. when the magnetic impurity is in the vicinity of a surface of a film or bulk material, we obtain that the MA has only diagonal non-zero elements, i.e. the energy difference of the 3/2 and 1/2 states defines the MA constant ($K_{L}=K_{1}$). Moreover, in case of cubic symmetry, when the magnetic impurity takes place in the bulk, $K_{1}$=0 and, therefore, $K_{L}$=0 as well in agreement with the statement that the $D^{3/2}$ ground state ($\Gamma_{8}$) remains degenerate in a cubic crystal field. This implies that the magnetic anisotropy should go to zero at the inner shells when increasing the size of nano-grains like in the bulk material (the characteristic energy scale of the MA goes by $N^{-3/2}$, see main paper). The calculated values in Fig. \ref{MAfig} are very fluctuating, which suggests that we are far from the bulk behavior. We also note that in case of ordered nano-clusters zero value was obtained for the central atom in agreement with the theoretical investigations based on symmetry analysis.

The parameters of the MA matrix can be calculated for each core site in a given nano-grain. The matrix elements are different for different sites even in the same shell, but symmetry relations were found between the sets of the MA parameters.
Let us enumerate the core sites by index $i=1...87$. We label expression (\ref{HL}) by a site index $i$, i.e. we now
introduce a set of the $H^{L}_{i}$-s. The parameter set $\left\lbrace K_{\mu}\right\rbrace$ will have a site index $i$, too. Let $i$ and $j$ be two sites in the same shell and let the two sites be connected
by the group element $g\in O_{h}$ where $g \mathbf{R}_{i} = \mathbf{R}_{j}$. So the
transformation rule for the matrices $H^{L}_{i}$ and $H^{L}_{j}$
can be written as
\begin{equation}
H^{L}_{j}=e^{i\varphi \mathbf{n}\cdot \mathbf{J}}H^{L}_{i}e^{-i\varphi \mathbf{n}\cdot \mathbf{J}}\;,
\label{szumtraf}
\end{equation}
where $\mathbf{J}=(J_{x},J_{y},J_{z})$ is the angular momentum-vector operator, while $g$ corresponds to the rotation around axis $\mathbf{n}$ with angle $\phi$. From Eq. (\ref{szumtraf}) we can derive relations between the anisotropy matrixelements of the different sites.

Motivated by the quadrupole-decomposition of the anisotropy
matrices, it seemed reasonable to
calculate the transformation rules for the coefficients in Eq. (5) of the main paper.
The quadrupolar operators form the basis of a real five-dimensional vector
space (parameter space), and the transformation matrices
between the parameter sets of the different sites $i$ and
$j$ are defined as
\begin{equation}
\mathbf{K}^{j}=\Gamma^{(i\rightarrow j)}(g)\mathbf{K}^{i} \;,
\end{equation}
where $\mathbf{K}^{i(j)}$ is a vector formed by the set of $\left\lbrace K_{\mu}^{i(j)}\right\rbrace$. The transformation matrices $\Gamma^{(i\rightarrow j)}(g)$ have a
very special structure, namely they are block-diagonal with two- and three-dimensional blocks: $E_{g}^{(i\rightarrow j)}(g)$ and $T_{2g}^{(i\rightarrow j)}(g)$ in order. The
five-dimensional parameter space is decomposed into two (dim=2+3) subspaces.

\begin{figure}[h!]
\begin{center}
\begin{tabular}
[c]{c}%
\includegraphics[width=0.8\textwidth, bb=  20 100 900 520]{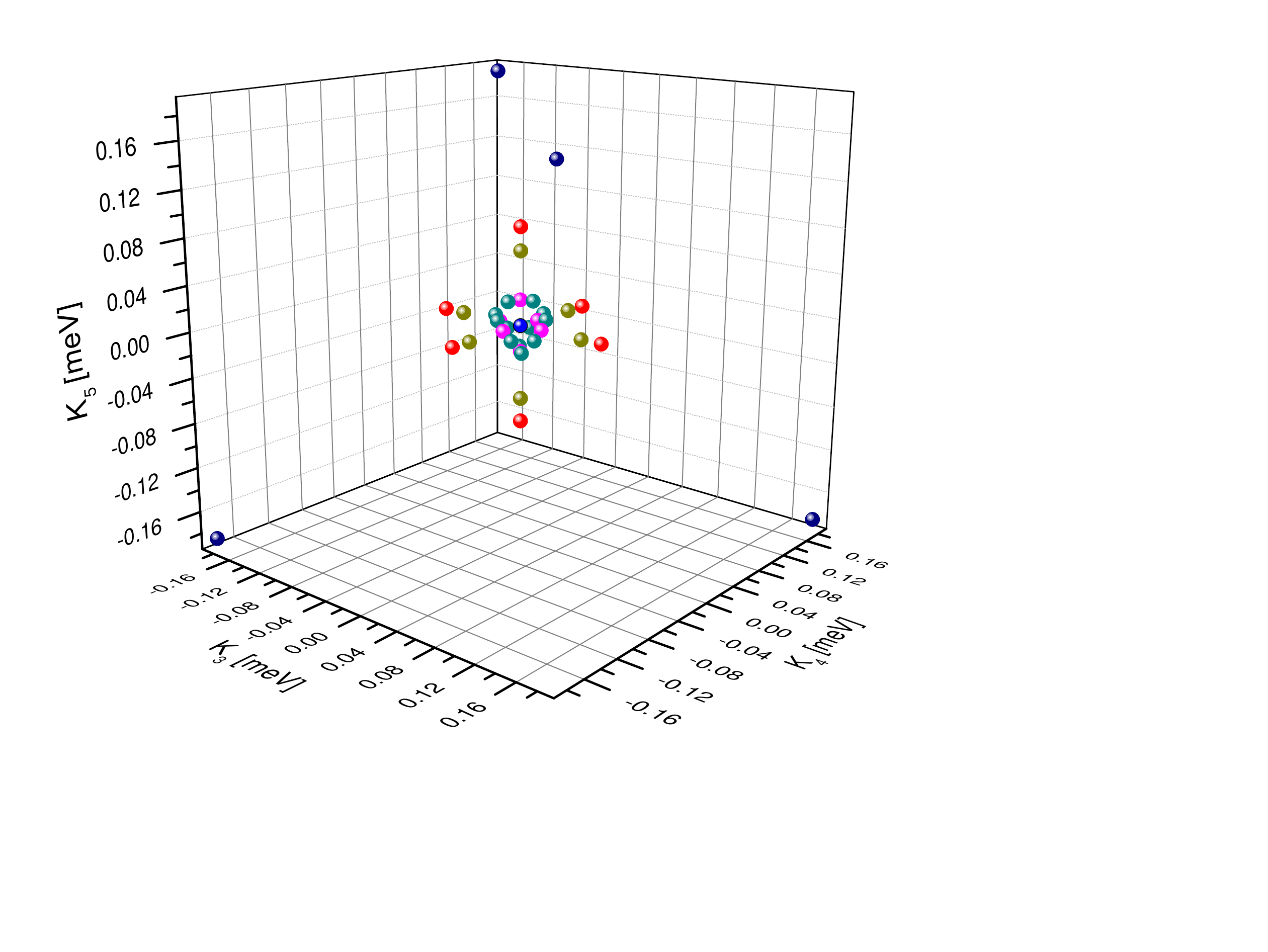}
\end{tabular}
\end{center}
\par
\vskip -0.5cm  \caption{(color online) The magnetic anisotropy parameters
in the $T_{2}$-space in case of an ordered nano-grain ($N$=225, $N_{c}$=87, $N_{S}$=1). The color coding is the same as in Fig. \ref{MAfig}.}%
\label{spd2}%
\end{figure}

\begin{figure}[h!]
\begin{center}
\begin{tabular}
[c]{c}%
\includegraphics[width=0.8\textwidth, bb=   20 0 890 350]{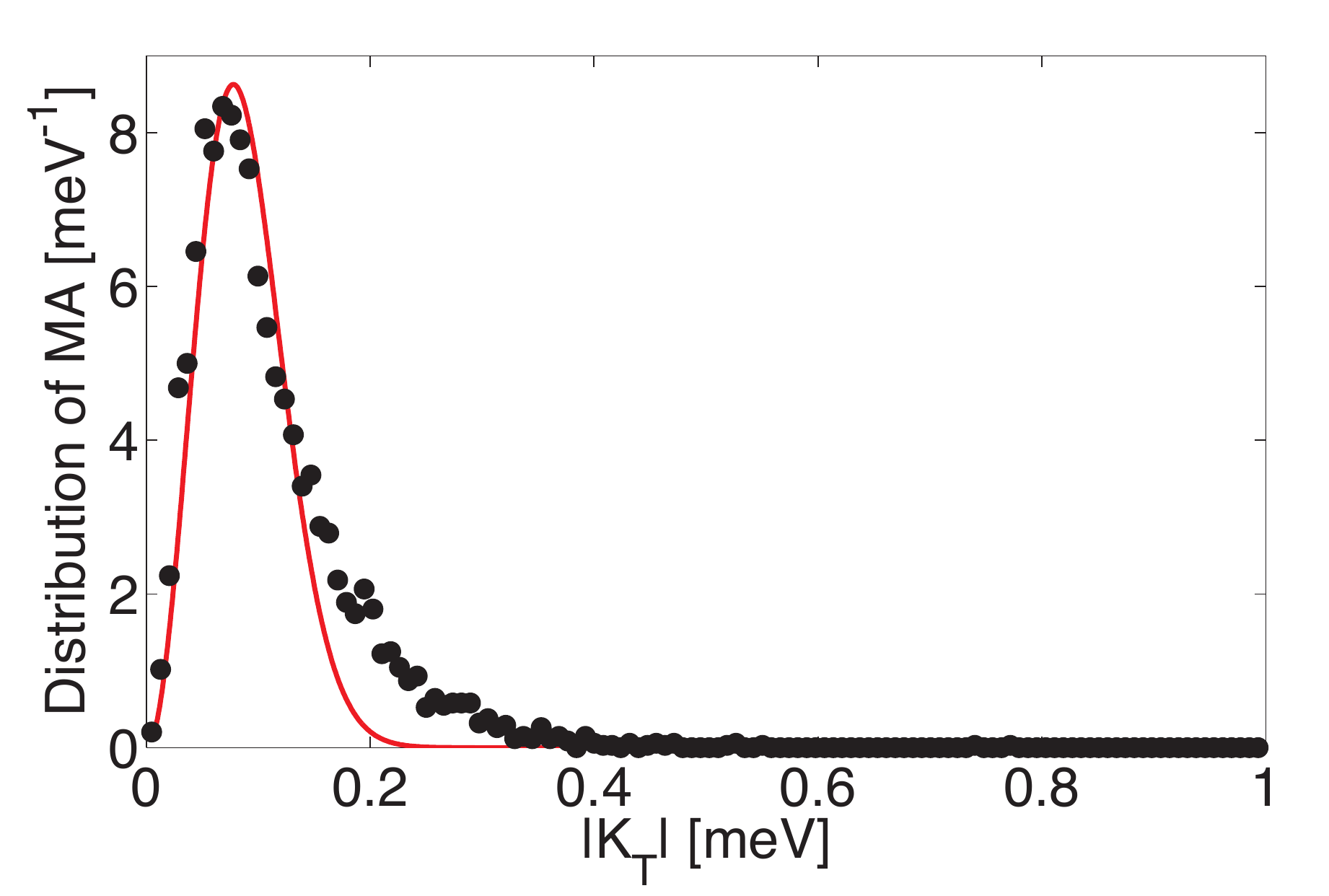}
\end{tabular}
\end{center}
\par
\vskip -0.5cm  \caption{(color online) Radial distribution of the magnetic anisotropy parameters (dots) in $T_{2}$-space in case of $N_{S}$=50 samples with $N$=225+25 atoms ($N_{c}$=87). The continuous line presents the prediction of the Gaussian Unitary Ensemble. The obtained MA energy scale: $\Delta_{T}$=0.13 meV.}%
\label{Tdisordered}%
\end{figure}

Whereas the two-dimensional $E$ parameter subspace has been discussed in the main paper, we focus here just on the three-dimensional $T_{2}$-space. We note that the $K_{3}$, $K_{4\text{ }}$ and $K_{5}$ parameters span the $T_{2}$ space being by about one order smaller in magnitude then the parameters on the $E$-plane. Fig. \ref{spd2} shows the calculated MA parameters in the $T_{2}$ space for the ordered nano-grain with 225 atoms (87 core atoms). We can identify the tetrahedral ordering pattern of the parameters in agreement of the structure of the $T_{2}$ transformation matrices. The ordered structure can be split even putting only one extra atom to the outmost shell of the spherical cluster.

Similarly to the case of $E$-plane, we examine the radial distribution of the MA parameters in the $T_{2}$ parameter space. In the simulations we dealt with the same 50 clusters as in case of Fig 3. in the main paper. The calculated $T_{2}$-space parameters are shown in Fig. \ref{Tdisordered}. The observed distribution agrees again with the prediction of Gaussian theory, similarly to Eq. (6) in the main paper. Vector $\mathbf{K}_{T}$ is defined as $\mathbf{K}_{T}\equiv(K_{3},K_{4},K_{5})$, and the $T_{2}$-space anisotropy scale $\Delta_T$ can be obtained as $\sqrt{\langle \mathbf{K}_T^2 \rangle}\equiv \Delta_T$. The $T_{2}$ data in Fig. \ref{Tdisordered} can be fitted by Gaussian unitary ensemble (GUE), see Appendix D for a definition. For the (normalized) distribution we obtained 5.103 for the $a_{2}$, and 1.986 for the $b_{2}$. However we found a better fit in the $E$-plane (it was fitted by GOE), there is no doubt that the power is $\beta$=2 for $T_{2}$ closer to the origin. $\Delta_{T}$ is really smaller than $\Delta_{E}$: $\Delta_{T}$=0.13 meV, while $\Delta_{E}$ was 2.37 meV.

\section{Appendix D - Level spacing distribution of the host-eigenenergies}

The spectral statistics of the Hamiltonian of the
disordered mesoscopic systems is usually analyzed in terms of random matrix theory (RMT). Here we analyze the energy level spacing distribution of the Hamiltonian Eq. (\ref{hamclus}) in the presence and in the absence of the host SO coupling. If $\xi=0$ in Eq. (\ref{hamclus}) then we speak of nonrelativistic grains, while by choosing the value $\xi=0.64$ eV relativistic grains are considered. 
The matrix $\hat{H}$ is an $M \times M$ matrix, where
$M=18\times N$ which has at most $M/2$ different eigenvalues $\left\{
\varepsilon_{i}\right\}  $ because of Kramers-degeneracy. The level spacings
$\left\{  s_{i}\right\}  $ are defined as
\begin{equation}
s_{i}=\varepsilon_{i+1}-\varepsilon_{i}\;,
\end{equation}
and $\varepsilon_{1}\leq\varepsilon_{2}\leq...\leq\varepsilon_{M/2}%
$. In case of disordered nano-grains the eigenvalues are strictly doubly degenerate,
so we always deal with $M/2$ different eigenvalues and, therefore, $M/2-1$ level spacing values $\left\{  s_{i}\right\}  $. In a given sample
there are $N_{S}\times M/2$ eigenvalues and $N_{S}\times(M/2-1)$ level
spacings. The distribution of $\left\{  \varepsilon_{i}\right\}$-s is nothing else then the DOS, Eq. (\ref{hostDOS}), while the distribution of $\left\{  s_{i}\right\}$-s  will be denoted by $p(s)$. We can introduce the universal $p(s)$ function as follows, 
\begin{equation}
p_{\beta}(x)=a_{\beta}x^{\beta}\exp\left(  -b_{\beta}x^{2}\right)  \;,
\label{wd2}%
\end{equation}
where the $\beta$ denotes the appropriate
Gaussian ensemble, and has a
crucial role in classification of different universal cases. $x=s/\Delta$, where $\Delta$ is the expected value of
the level spacings. In case of $\beta=1$ one can speak about the Gaussian Orthogonal Ensemble (GOE), the $\beta=2$ case defines the Gaussian Unitary Ensemble (GUE), while $\beta=4$ corresponds to the Gaussian
Symplectic Ensemble (GSE). If the time reversal symmetry of a
system is broken, e.g. there is an external magnetic field, then $\beta=2$,
and the GUE can be the proper ensemble. In the
presence of time reversal symmetry for particles with integer spins, the GOE is the appropriate ensemble, i.e., an orthogonal matrix diagonalizes the Hamiltonian. Without host SO
interaction the distribution of the level spacings is described by GOE. In
case of particles with half integer spins, and in the presence of the
host SO coupling, one has to deal with GSE \cite{mehta}. We note that the constraints
\begin{equation}
\int_{0}^{\infty}p_{\beta}(x)dx=1 \text{ and } \int_{0}^{\infty}xp_{\beta}(x)dx=1
\end{equation}
imply that
\begin{equation}
a_{\beta}=2\frac{\Gamma^{\beta+1}((\beta+2)/2)}{\Gamma^{\beta+1}((\beta
+1)/2)}\text{ and }b_{\beta}=\frac{\Gamma^{2}((\beta+2)/2)}{\Gamma^{2}%
((\beta+1)/2)}\;, \label{alphabetha}%
\end{equation}
where $\Gamma$ is the gamma function, $\Gamma(n)=(n-1)!$, and
\begin{align}
a_{1}  &  =\frac{\pi}{2}\approx1.571,\;b_{1}=\frac{\pi}{4}\approx
0.785\; \text{(GOE)}\;,\nonumber\\
\;\text{ }a_{2}  &  =\frac{32}{\pi^{2}}\approx3.242,\;b_{2}=\frac{4}{\pi
}\approx1.273\;\text{(GUE)}\;,\nonumber\\
a_{4}  &  =\frac{262144}{729\pi^{3}}\approx11.597,\;b_{4}=\frac{64}{9\pi
}\approx2.264\;\text{(GSE)}\;. \label{param}%
\end{align}

The calculated density of states, $n(\varepsilon)$, in case of a nano-grain host of $N$=225 Au atoms
can be seen in Fig. \ref{spd1}. The large DOS arises from the '$d$-band' around $3$ eV. Let us denote the integrated DOS by $N(\varepsilon)$. Calculating $N(\varepsilon)$ we observed two energy ranges, $\mathcal{E}_{d}$ and $\mathcal{E}%
_{s}$, where the integrated DOS is a nearly linear function and it can
be supposed, that $n(\varepsilon)$ is constant in these ranges, at least on
averages. The average level spacing can then be defined as,%
\begin{equation}
\Delta_{i}=\frac{1}{(M/2)n(\mathcal{E}_{i})}\\;\;\;\;(i=d,s). \label{delta}%
\end{equation}
In case of $d$-band between 3 and 4.5 eV ($\mathcal{E}_{d}$) a
good statistical analysis can be done because of the large values of $n(\mathcal{E}_{d})$.
The other energy range, %
$\mathcal{E}_{s}$ is chosen from 6 to 14 eV, where
the dominant component of the DOS is the $s$-type states.
The value of $n(\mathcal{E}_{s})$ is smaller than $n(\mathcal{E}_{d})$, 
therefore we cannot do a satisfactory statistical analysis. Importantly, the range $\mathcal{E}_{s}$ contains the Fermi energy.

Kubo-relation says that the expected value of the level spacings
is inverse proportional to $L^{d}$, where $L$ stands for the linear size of a $d$
dimensional system. It follows that%
\begin{equation}
\Delta_{N,i}\sim\frac{1}{N}\; ; \;\;\;\;(i=d,s)\;. \label{kubo2}%
\end{equation}
Reassuringly, the ratios of, say, $\Delta_{393,d}=2.337$ meV and $\Delta
_{273,d}=3.667$ meV agree with the value of$\ \frac{237}{393}$ within an
accuracy of $10$ $\%$ in case of nonrelativistic particles and the ratio of
$\Delta_{237,d}=3.983$ meV \'{e}s $\Delta_{165,d}=5.469$ meV agree with
$\frac{165}{237}$ within an accuracy of $5$ $\%$ in the presence of host SO coupling.

\begin{figure}[h!]
\begin{center}
\begin{tabular}
[c]{c}%
\includegraphics[width=0.8\textwidth, bb=  25 0 900 350]{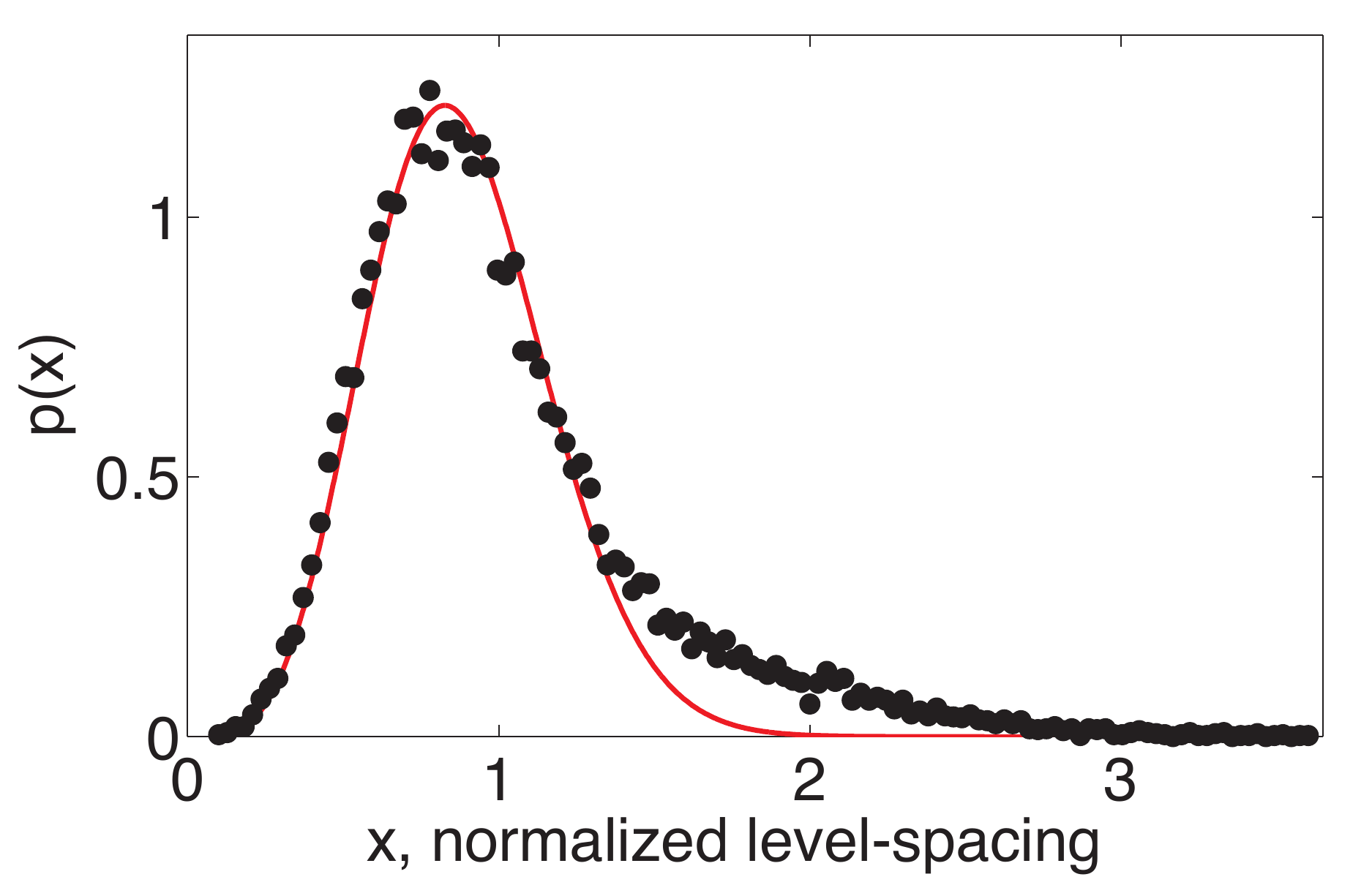}
\end{tabular}
\end{center}
\par
\vskip -0.5cm  \caption{(color online) The level spacing distribution for
$N_{S}$=50 disordered nano-grains with $N$=225+25 atoms in the $d$-band. The average level spacing is $\Delta=3.9$ meV.
The data are well fitted by GSE function, $a_{4}$=19.13, $b_{4}$=2.867 with 95 \% confidence bound, 0.965, and are in agreement with the experiments \cite{levelexp}.}%
\label{RMT}%
\end{figure}

The distribution of the level spacings is shown in Fig.
\ref{RMT}. We dealt with 225+25 atoms in a given cluster and chose randomly 25 sites from the next 96 sites at the 12NN, 12NNa and 13NNb shells (see Table \ref{tablegrain}), which creates $C^{96}_{25}$ possible configurations. In our simulation $N_{S}=50$ relativistic samples were calculated. The data are well fitted by GSE: $a_{4}$=19.13, $b_{4}$=2.867 with 95 \% confidence bounds. 

It should be noted that the largest deviation of the calculated data from the universal $p(x)$ function can be observed in the tail region. The deviation of the host level spacing statistics (fat tail) from the prediction of RMT together with the anomalies observed in anisotropy parameter space (angle correlations) strongly suggest that the effect of the cubic symmetry of the underlying lattice is not negligible compared with total randomness. Most importantly, as we mentioned in the main paper, the obtained level spacing distribution in Fig. \ref{RMT} is in agreement with the experiments \cite{levelexp}.

\section{Appendix E - Magnetic excitation spectrum of impurities in nano-grains}

Here we calculate the excitation spectra of magnetic impurities embedded in random nano-grains. The LSO MA energy can be calculated due to the effective Hamiltonian $H^{L}$, see Eqs. (2) and (5) in the main paper. To describe the spectra of an arbitrarily oriented sample we rotate $H^{L}$ by an angle $\varphi$ about the unit vector $\mathbf{n}$. Then the Hamiltonian becomes
\begin{eqnarray}
H^{L}_{\mathbf{n},\varphi}= e^{i\varphi \mathbf{n}\cdot \mathbf{J}}H^{L}e^{-i\varphi \mathbf{n}\cdot \mathbf{J}}
\end{eqnarray}
where $\mathbf{J}=(J_x,J_y,J_z)$ is the angular momentum-vector operator. We solved the eigenproblem of this Hamiltonian $H^{L}_{\mathbf{n},\varphi}|i\rangle=E_i|i\rangle$  ($i=1,2,3,4$). The Fermi's golden rule for the intensities is given as
\begin{eqnarray}
S(\omega)\propto\sum_{j>i}\delta(\omega-(E_j-E_i)/\hbar)|\langle j|J_x|i\rangle|^2p_i,
\label{esrspect}
\end{eqnarray}
where $p_i=\frac{e^{-\beta E_i}}{Z}$ is the Boltzmann weight of the $i$th eigenstate, and $Z=\sum_i e^{-\beta E_i}$ is the partition function with $\beta=1/k_BT$ the inverse temperature.

We simulated the excitation spectra of disordered nano-grains with $N$=225+40 atoms according to the following procedure: we generated $N_{S}$=50 nano-grains, and for each such grain we placed one magnetic ion on the different core sites. There are 87 such sites, which means we generated 4350 different samples.  From these 4350 nano-grains we have randomly chosen 100 clusters, and rotated them with angles $\varphi$ uniformly distributed in the interval $\left[0,\pi\right]$ and chose the vector $\mathbf{n}$ uniformly distributed on the unit sphere. The Dirac-delta in Eq. (\ref{esrspect}) was replaced by an (unnormalized) Lorentzian with half-width $\Gamma=0.1$ THz for the absorption spectra (for better comparison with experiments we give the frequencies in THz). The temperature was set to $T=4.2 \ K$. All these 100 spectra were added up, and the procedure was repeated ten times. The obtained results are shown in Fig. 4 of the main paper. 

\section{Appendix F - Schottky anomalous heat capacity for nano-grains with magnetic impurities}

\begin{figure}[h!]
\begin{center}
\begin{tabular}
[c]{c}%
\includegraphics[width=0.8\textwidth, bb=  10 0 1250 580]{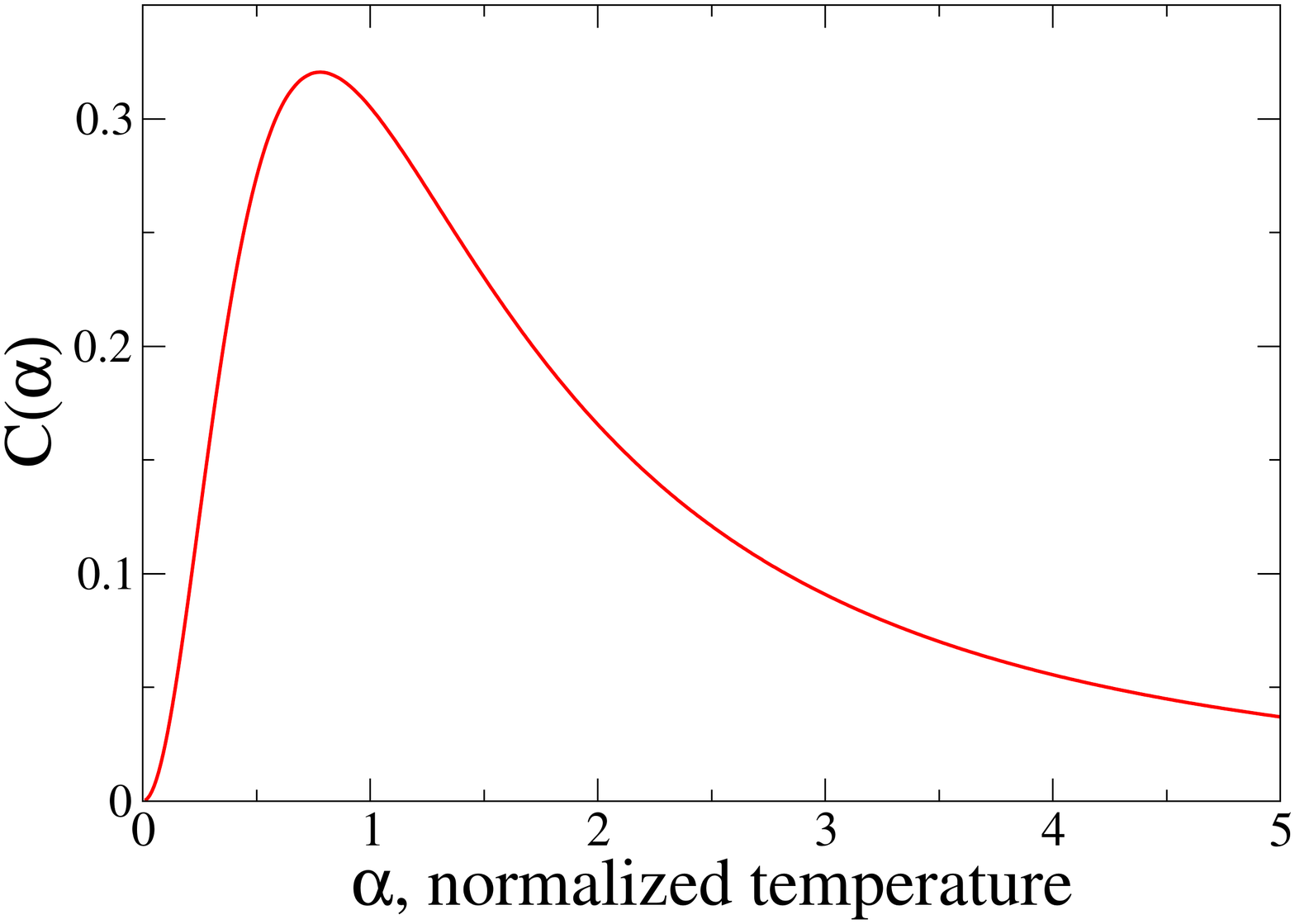}
\end{tabular}
\end{center}
\par
\vskip -0.5cm  \caption{(color online) Specific heat (in units of $k_B$) of nano-grains with magnetic impurities, obtained from the  numerical integration in Eq. (\ref{SH4}) as a function of the normalized temperature, $\alpha=k_{B}T/ \Delta_{E} $. This curve is universal for all disordered nano-grains, where the MA  constants follow a GOE-type of distribution, Eq. (\ref{OMdef2}). The grain-specific informations are hidden in the parameter $\Delta_E$. 
}
\label{speciheat}%
\end{figure}

Here we derive the specific heat (heat capacity) of the nanoballs with magnetic impurities. 
We have seen that the MA matrix has two Kramers degenerate eigenstates with the energy splitting equal to $2K_{L}$, where $K_{L}$ is the MA constant, see Eqs. (3-4) in the main paper. The canonical partition function for a single impurity can be written as
\begin{equation}
 Z=2(1+e^{-2K_{L}\beta})\;,
\label{Zdef}
\end{equation}
where $\beta$=$(k_{B}T)^{-1}$ is the inverse temperature. The thermodynamic value of the energy is defined as
 \begin{equation}
 E(T) =- \frac{\partial \ln Z} {\partial \beta} = \frac{2K_{L}}{1+e^{2K_{L}\beta}}\;.
\label{Edef}
\end{equation}
Fig. 2 in the main paper demonstrates that the main (E-plane) contribution to the MA constant follows a GOE-type random distribution,
\begin{equation}
p(K_L)=a K_L e^{-b K_L^{2}}\;.
\label{OMdef}
\end{equation}
with $\Delta_{E}=\frac{1}{\sqrt{b}}$, see Eq. (6) in the main paper. Note that $a$=$2b$ because $p(\varepsilon)$ is normalized to unity. Hence Eq. (\ref{OMdef}) can be written as
\begin{equation}
p(K_L)=\frac{2}{\Delta_{E}^{2}} K_L e^{ -\frac {K_L^{2}} {\Delta_{E}^{2}}}\;.
\label{OMdef2}
\end{equation}
By using Eq. (\ref{Edef}), we calculate the average of the energy over this distribution,
\begin{equation}
 \langle E(T) \rangle=\int_{0}^{\infty}\frac{2K_L}{1+e^{2 K_L \beta}} p(K_L)dK_L \;.
\label{Edef2}
\end{equation} 
Inserting Eq.~(\ref{OMdef}) into Eq. (\ref{Edef2}), and introducing the new variable $ x \equiv K_L \beta $, we obtain for the expectation value of the energy,
\begin{equation}
 \langle E (T)\rangle=\frac{4 k_{B}^{3}T^{3}}{\Delta_{E}^{2}} \int_{0}^{\infty}  \frac{x^{2}  e^{-\frac{k_{B}^{2}T^{2}  x^{2}}{\Delta_{E}^{2}}}} {1+e^{2x}} dx \;
\label{E2}
\end{equation}
or, alternatively,
\begin{equation}
 \langle E(\alpha) \rangle=4\Delta_{E} \alpha^{3} \int_{0}^{\infty}  \frac{x^{2}  e^{- \alpha^{2}  x^{2}}} {1+e^{2x}} dx \;,
\label{E3}
\end{equation}
where we introduced the normalized temperature $ \alpha \equiv \frac{k_{B}T}{\Delta_{E}} $. The averaged specific heat,
\begin{equation}
\langle C(T) \rangle = \frac{d \langle E(T) \rangle } {d T}\;,
\label{Cdef}
\end{equation}
can then be  expressed as a function of $\alpha$ as
\begin{equation}
\langle C (\alpha) \rangle= k_{B} \alpha^{2}\int_{0}^{\infty} \left(12x^{2}-8 \alpha^{2} x^{4} \right)\frac{ e^{- \alpha^{2}  x^{2}}} {1+e^{2x}} dx\;.
\label{SH4}
\end{equation}
To get the temperature dependent specific heat, the integration in Eq. (\ref{SH4}) has to be performed numerically for different values of $\alpha$. The result of this procedure can be seen in Fig. \ref{speciheat}. Evidently, the random (GOE) distribution of the MA constants induces a Shottky peak in the specific heat at $\alpha \approx 0.78$, i.e. at $T_m\approx 0.78\; \Delta_E $. It can also be noted that $C(\alpha)$ is proportional to $\alpha^{2}$, thus $C(T)\sim T^2/\Delta_E^2$ for temperatures, $T \ll T_m$.

